\documentclass[10pt,journal,compsoc]{IEEEtran}

\usepackage[pdftex]{graphicx}
\usepackage{amsmath}
\ifCLASSOPTIONcompsoc
  \usepackage[caption=false,font=footnotesize,labelfont=sf,textfont=sf]{subfig}
\else
  \usepackage[caption=false,font=footnotesize]{subfig}
\fi

\usepackage{pdfcomment}
\usepackage{enumitem}
\usepackage{footnote}
\usepackage[bottom]{footmisc}
\usepackage{tabularx}
\usepackage[linesnumbered,ruled]{algorithm2e}
\usepackage{multirow}
\usepackage{color}

\usepackage{algpseudocode}
\algnewcommand{\Or}{\textbf{ or }}
\algnewcommand{\And}{\textbf{ and }}

\begin{document}

\title{ENORM: A Framework For \textbf{E}dge \textbf{NO}de \textbf{R}esource \textbf{M}anagement}

\author{Nan~Wang,
        Blesson~Varghese,
        Michail~Matthaiou~
        and~Dimitrios S. Nikolopoulos

\IEEEcompsocitemizethanks{\IEEEcompsocthanksitem N. Wang, B. Varghese, M. Matthaiou and D. S. Nikolopoulos are with Queen's University Belfast, UK.\protect\\
E-mail: \{nwang03, b.varghese, m.matthaiou, d.nikolopoulos\}@qub.ac.uk
}
\thanks{Manuscript received January 6, 2017; accepted September 12, 2017.}}

\markboth{IEEE Transactions on Services Computing,~Vol.~X, No.~Y, September~2017}%
{Wang \MakeLowercase{\textit{et al.}}: ENORM: A Framework for Edge NOde Resource Management}

\IEEEtitleabstractindextext{%
\begin{abstract}
Current computing techniques using the cloud as a centralised server will become untenable as billions of devices get connected to the Internet. This raises the need for fog computing, which leverages computing at the edge of the network on nodes, such as routers, base stations and switches, along with the cloud. However, to realise fog computing the challenge of managing edge nodes will need to be addressed. This paper is motivated to address the resource management challenge. We develop the first framework to manage edge nodes, namely the \textit{E}dge \textit{NO}de \textit{R}esource \textit{M}anagement (ENORM) framework. {\color{black} Mechanisms} for provisioning and auto-scaling edge node resources are proposed. The feasibility of the framework is demonstrated on a Pok\'eMon Go-like online game use-case. The benefits of using ENORM are observed by reduced application latency between 20\%-80\% and reduced data transfer and communication frequency between the edge node and the cloud by {\color{black} up to 95\%}. These results highlight the potential of fog computing for improving the quality of service and experience. 
\end{abstract}

}

\maketitle

\IEEEdisplaynontitleabstractindextext

\begin{figure}
  \centering
    \includegraphics[width=0.48\textwidth]{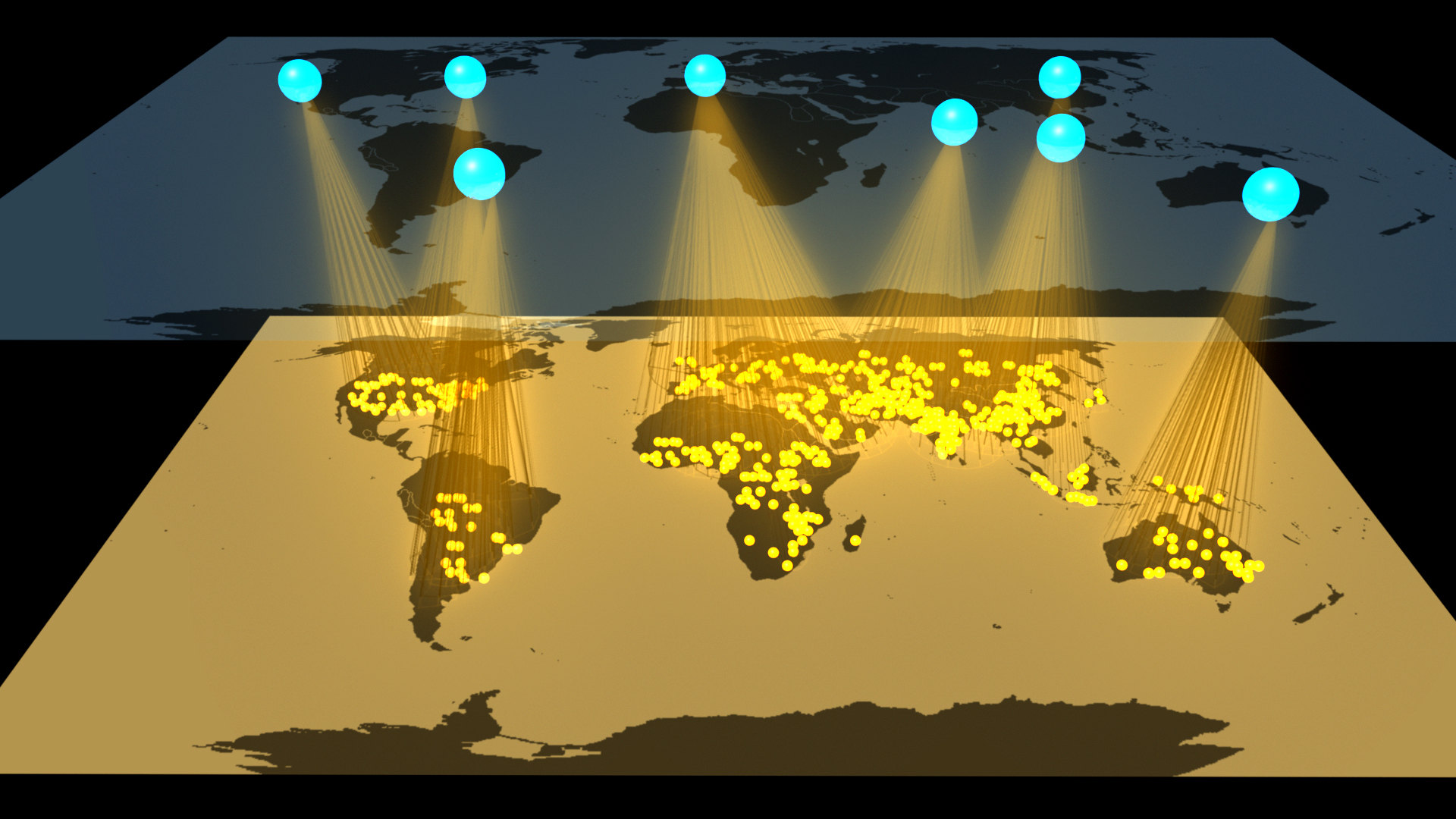}
  \caption{Existing computing model that employs the cloud. Blue dots are example locations of cloud data centers and yellow dots show devices that interact with servers in the data center.}
  \label{fig:introCloudOnly}
\end{figure}

\IEEEraisesectionheading{
\section{Introduction}
\label{sec:introduction}}
Currently, applications that are hosted on cloud servers, typically provide services from a cloud data center as shown in Figure~\ref{fig:introCloudOnly}. Some of these servers are replicated across multiple data centers located in different geographical regions for reducing the workload
and improving the Quality-of-Service (QoS). However, this will not be suitable in the future given that over 25 billion devices are estimated to be added to the network by 2020\footnote{\url{http://www.gartner.com/newsroom/id/2636073}}.
The QoS will degrade as more devices need to be catered for by existing communication and computing infrastructure~\cite{paper1}. 
Three problems will need to be addressed to tackle this. They are minimising communication latency between users and computing devices and reducing both data traffic to the cloud and communication frequency between devices and the cloud. 

\begin{figure}
  \centering
    \includegraphics[width=0.48\textwidth]{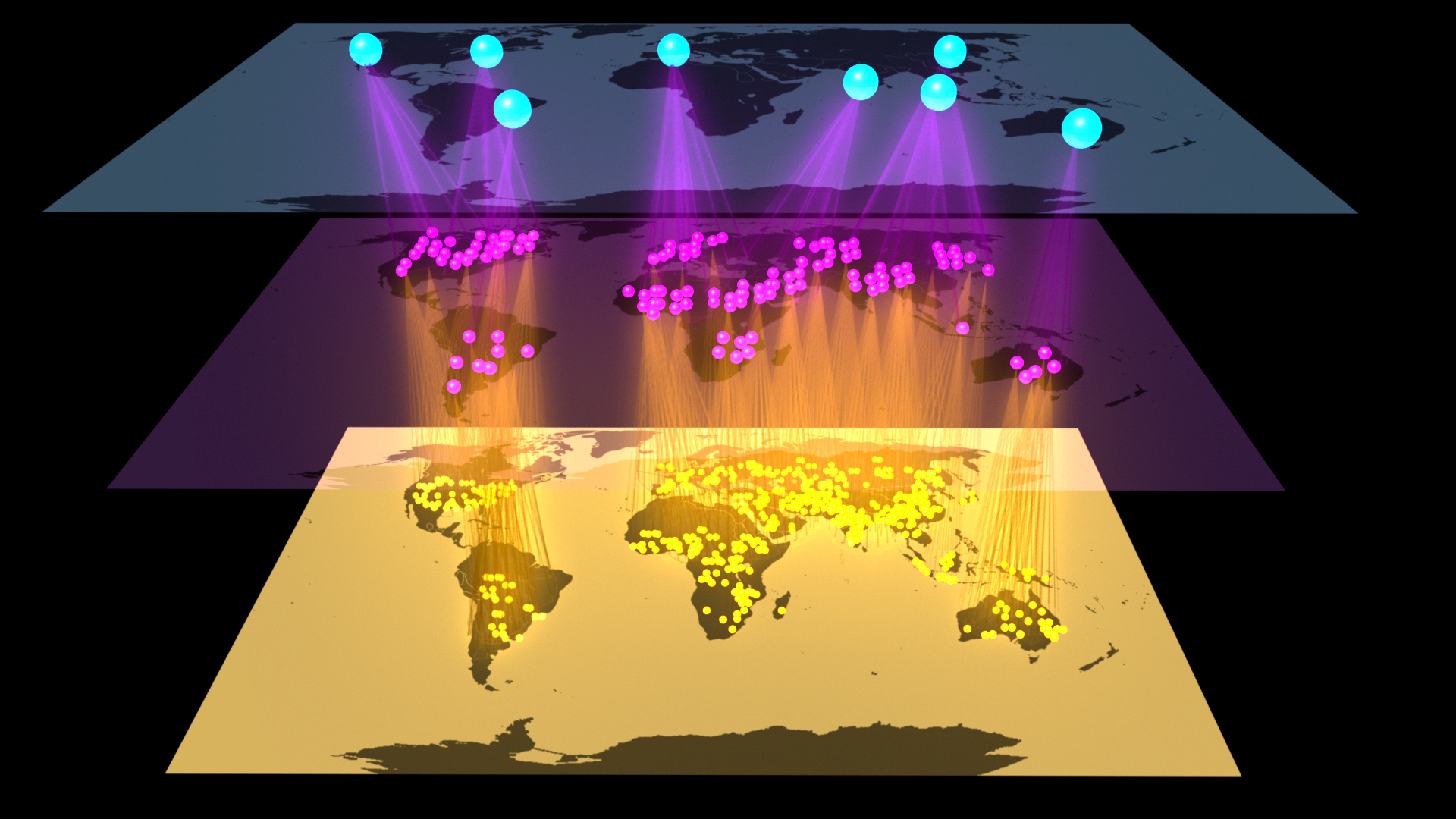}
  \caption{The fog computing model. While the blue dots show example locations of cloud data centers, the purple dots highlight the larger volume of edge nodes that are present in the network. The user devices communicate more frequently with the edge nodes, which in turn interact less frequently with the cloud servers.}
  \label{fig:introCloudEdge}
\end{figure}

In this paper, we prototype a \textit{`Fog Computing'}~\cite{fogcomputing-00,edgecomputing-00} framework to bring computing towards the edge of the network, which is closer to the user devices, so that the latency of communication, data traffic to the cloud and frequency of communication between user devices and the cloud can be reduced. This is shown in Figure~\ref{fig:introCloudEdge}. In contrast to the two-tier cloud computing architecture, in which a user interacts only with the cloud, the fog computing architecture is at least a three-tier architecture. An additional edge node {\color{black} (for example, mobile base stations, routers, and switches)} layer is inserted between the user and the cloud. 

Resource management at the edge of the network will play an important role in realising fog computing. 
However, this is challenging in a number of ways. 
{\color{black} We have identified three problems that will need to be solved in this space.}

The {\color{black}first problem} is provisioning edge nodes for executing workloads {\color{black} offloaded} from the cloud. While there exists a marketplace for executing workloads on the cloud, such a marketplace is unavailable for edge nodes. This makes it challenging to provision an edge node since directives and standard protocols for initialising services on a potential edge node and for communicating between the cloud and edge nodes are not fully known and developed. 

The {\color{black}second problem} is deploying workloads on edge nodes. The following two important questions must be considered: (i) How to deploy a workload on the edge node? On the cloud, Virtual Machines (VMs) are usually employed to execute a workload. However, VMs are less likely to be suitable on edge nodes given the availability of limited hardware resources. 
(ii) How much of the workload can be deployed on the edge node? Large workloads are easily executed on the cloud given the access to a large amount of resources.
{\color{black} It is challenging to decide appropriate workloads to execute on an edge node given its limited hardware resources and varying computation availability introduced by its primary service. For example, the primary service of a Wi-Fi access point is to route Internet or mobile traffic, which we refer to as the \textit{basic service}.}

The {\color{black}third problem is managing resources on edge nodes}. 
{\color{black} Due to its basic service,} workloads offloaded to an edge node are of secondary priority, since the primary service cannot be compromised. 
{\color{black} This makes resource management on edge nodes challenging because} the resources allocated for the workload will have to dynamically scale (or auto-scale) given that there are limited hardware resources available on the edge node when compared to a cloud server. 
Additionally, the allocation of resources to host multiple tenants will need to be considered.

Existing resource management frameworks in distributed computing are suitable in the context of clusters~\cite{clustermanagement-2}
and clouds~\cite{cloudmanagement-2},
but do not consider the edge of the network.
In this paper, we aim to tackle the above three problems by proposing and developing ENORM, an \textit{E}dge \textit{NO}de \textit{R}esource \textit{M}anagement framework that integrates the edge of the network in the computing ecosystem to realise fog computing. 
We firstly propose a novel provisioning {\color{black}and deployment} mechanism 
to integrate an edge node with a cloud server. 
Contrary to cloud-based deployments, our technique is developed {\color{black} for resource limited environments. Consequently, our mechanism only requires simple implementation.} 
Secondly, an auto-scaling mechanism to dynamically manage edge resources is proposed. This mechanism is essential to meet service objectives for improving the QoS and to safely use edge nodes without affecting other workloads that are executed on them. 

The feasibility of ENORM is validated on a Pok\'eMon Go-like location-aware {\color{black} and latency-sensitive} online game use-case. The key result is that there are significant benefits in improving the QoS of a large number of users in a given location by employing ENORM for achieving the fog computing based use-case. The low overheads of the framework not only indicate that fog computing is feasible, but also demonstrate the lightweight nature of the mechanisms developed within ENORM. When compared to a cloud-only model, the application latency is reduced between 20\%-80\%. Similarly, the data traffic and the communication frequency between the edge node and the cloud server are both reduced by up to {\color{black} 95\%}. All these results showcase the benefit of integrating the edge in the computing ecosystem with the cloud in fog computing. 

The research contributions of this paper are firstly the integration of the edge of the network in the computing ecosystem through the development of a provisioning mechanism that facilitates communication between edge nodes and the cloud. Currently, there are no distributed computing frameworks that can manage edge node resources in the computing ecosystem. ENORM on the other hand can provision edge nodes for offloaded workloads from the cloud. Secondly, ENORM incorporates a low overhead and dynamic auto-scaling mechanism to add or remove resources for efficiently handling the workloads on the edge node. We note that using ENORM over 16,000 users can be serviced by the edge node by only incurring an additional 5-second overhead for auto-scaling. 

The remainder of this paper is organised as follows. 
Section~\ref{sec:framework} presents ENORM for managing resources of edge nodes. 
Section~\ref{sec:resourcemanagement} considers provisioning and auto-scaling in ENORM.
Section~\ref{sec:usecase} demonstrates an online game use-case employed to test the feasibility of fog computing using ENORM. 
Section~\ref{sec:evaluation} evaluates the overheads and performance gain of using ENORM. 
Section~\ref{sec:relatedwork} presents related work.
Section~\ref{sec:conclusions} concludes this paper.

\section{The ENORM Framework}
\label{sec:framework}
\begin{figure}
  \centering
    \includegraphics[width=0.5\textwidth]{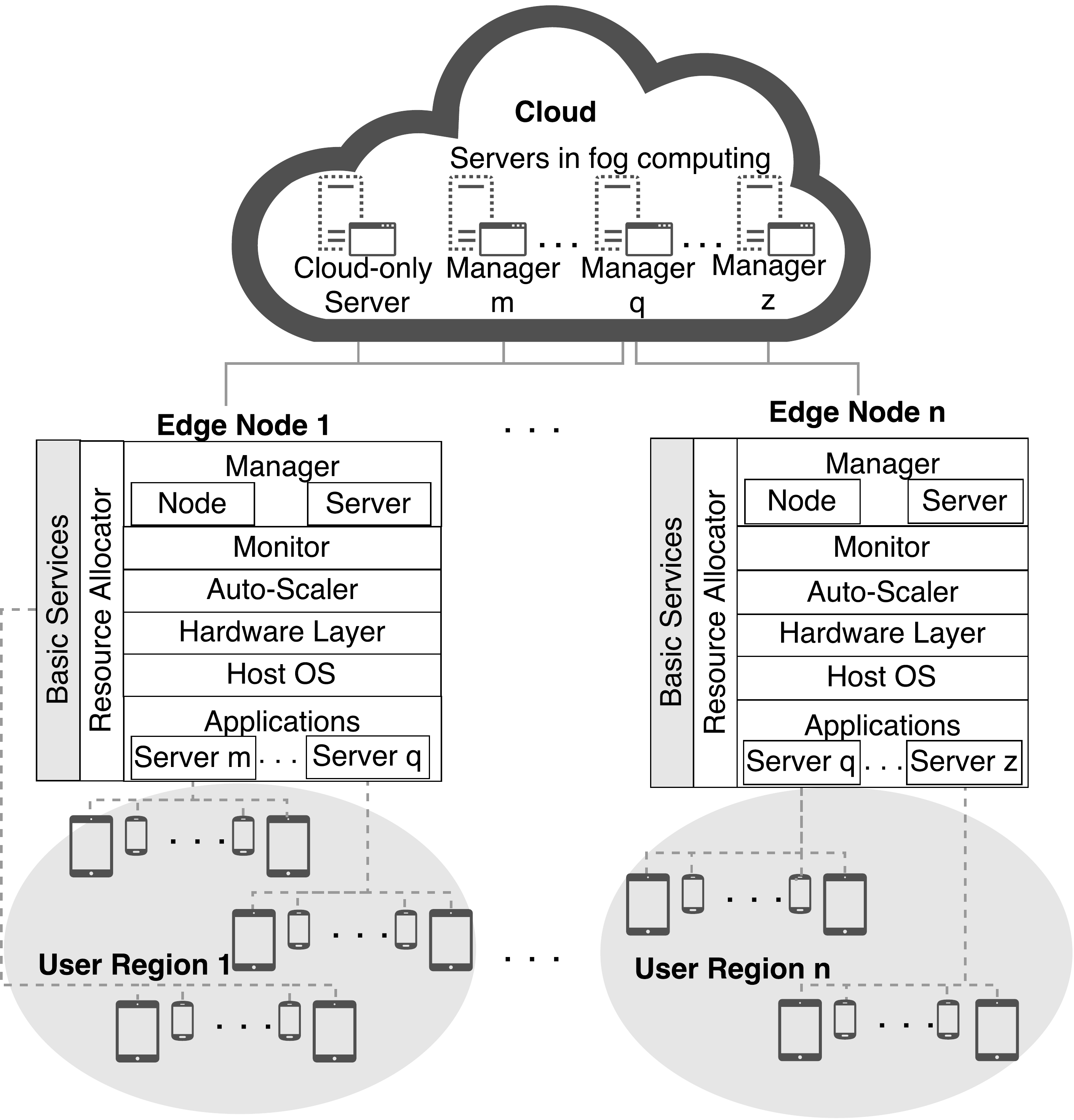}
  \caption{Architecture of ENORM in the fog computing ecosystem. Typically, using the cloud-only model user devices connect to a cloud server through the basic service offered by a traffic routing node. In ENORM, in addition to the basic services, user devices are serviced by the edge nodes that host servers offloaded from cloud servers.}
  \label{fig:systemarchitecture}
\end{figure}

To {\color{black} solve the three challenging problems presented before} in realising fog computing, we propose a resource management framework, referred to as `\textbf{ENORM}: \textbf{E}dge \textbf{NO}de \textbf{R}esource \textbf{M}anagement'. 
The architecture we propose for ENORM works across three-tiers as shown in Figure~\ref{fig:systemarchitecture}. The \textit{top tier} is the cloud tier, where application servers are hosted. In the cloud-only execution model, connections from user devices are established with the servers located in one or more data centers.

To enable the use of the cloud in conjunction with edge nodes, our framework deploys a \textit{cloud server manager} on each application server. This server (i) communicates with potential edge nodes requesting computing services, and (ii) deploys partitioned servers on the edge node, and (iii) receives updates from the edge node to update the global view of the application server on the cloud. 

The \textit{bottom tier} is the user device tier. Multiple devices, such as smartphones, wearables and gadgets, are connected to the cloud application servers. In the cloud-only execution model, the devices connect to the application servers through traffic routing nodes. The basic services of the mobile base stations in the cellular network or routers in the wireless local network are used to communicate to the centralised cloud server. However, using ENORM, the computing capabilities of the edge nodes are made use of. 

The \textit{middle tier} is the edge node tier. This could be multiple layers of edge nodes that scale both horizontally and vertically. In this paper, the focus is on a single edge node (this is the first framework for fog computing resource management and multiple nodes will be considered in the future). As seen in Figure~\ref{fig:systemarchitecture}, the servers making use of the edge communicate with on-demand edge servers to support users in regions covered by each edge node. For example, users start an application (App $q$, which is hosted by Server $q$) in multiple regions covered by edge nodes (Edge Node $1$ and $n$). Connections are established with the respective application servers on the cloud. A partitioned application server of App $q$ is deployed to the edge nodes (the details of how the edge node is provisioned and connections are established will be presented in Section~\ref{sec:resourcemanagement}). 

The partitioned server on the edge node is different from the cloud application server in that localised data relevant to the users covered by the edge node is maintained. The global view is maintained on the cloud server and the edge node updates the cloud server. When edge nodes cannot provide computing services (for example, if the edge node is overloaded during peak hours and there are no spare cycles) or the edge nodes cannot improve the QoS of the application, then the deployed edge server will be terminated and users connect to the cloud application server as in the cloud-only execution model. 

Our proposed ENORM framework, uses the following five components on the edge node. They are:   

(i) \textit{Resource Allocator}: In providing computing as a service, the basic service of an edge node cannot be compromised. This involves prioritising the basic service over any offloaded workloads from cloud servers. It is therefore essential to know the free resources that are available on the edge node.
The resource allocator keeps track of the available CPU cores and memory. 

(ii) \textit{Edge Manager}: The Edge Manager comprises two sub-components, namely the Node Manager and the Server Manager. 
The node manager deals with the requests that are obtained by the server manager from a cloud server. When a request is made from the cloud manager, a decision on whether to accept the request is made by the node manager (this decision is based on whether {\color{black} there are free resources on the edge node and whether} the priority of the requesting application is higher than or equal to that of an executing server on the edge). The response is sent back to the requesting cloud server. Once a request is accepted, the server manager initialises a container, allocates necessary ports for communicating and updates firewall settings. 

In our framework, Linux containers are employed to provide applications isolation on the edge node through virtualisation. Our rationale for choosing container virtualisation technology instead of VMs is due to the reduced boot up times and enhanced isolation provided by the former~\cite{linuxcontainers-1}. Moreover, on limited hardware platforms, such as edge nodes, containers are appropriate given that they are relatively lightweight and have low overheads.

(iii) \textit{Monitor}: A number of metrics related to each application edge server is monitored periodically. Our monitor tracks communication latency (using standard commands such as ping and assuming that the user devices have static IP addresses) and computing latency (obtained from time-stamps on server logs). These metrics are employed by the auto-scaler on whether more resources need to be allocated or existing resources need to be removed from an application server. For example, the edge server performance is monitored in terms of round-trip application latency and hardware utilisation of CPU and memory. 

(iv) \textit{Auto-scaler}: Based on the metrics obtained from resource allocator and monitor, the auto-scaler dynamically allocates/de-allocates hardware resources to the containers executing application servers (considered in Section~\ref{sec:resourcemanagement}). Dynamic allocation of resources is required to (i) ensure that the basic service has sufficient resources such that there is no overload due to additional application servers, and (ii) modify allocated resources to accommodate more users or new application servers on the edge node. If an application server cannot obtain resources or even if obtained could not improve performance, then the edge manager terminates the edge server. This is considered further in Section~\ref{autoscaling}.

(v) \textit{Application Edge Server}: The partitioned server from the cloud is hosted on the edge node. This server interacts with user devices before forwarding data to the cloud. 

\section{Resource Management in ENORM}
\label{sec:resourcemanagement}
\begin{table*}[t]
\centering
\caption{Notation used in the proposed resource management mechanisms}
\begin{tabular}{c  l  c} \hline
\textbf{Parameter} & \textbf{Description} & \textbf{Source} \\ \hline
\(service\) & Flag on the availability of the services on an edge node & \multirow{ 5}{*}{Edge manager}\\ 
\(S\) & A set of $n$ edge servers hosted on $n$ containers in an edge node, ordered by $Pri$. $s_i \in S, i=1,..., n$ & \\ 
\(r^u\) & One unit of CPU and memory &\\
\(Prt\) & A list of all available ports on an edge node &\\
\(termType\) & Flag for 'single' or 'multiple' termination of containers on an edge node &\\
\hline
\(R\) & Free CPU and memory on the edge node available for $S$ & Resource allocator\\ \hline
\(Pri_i\) & Priority of $s_i$ & \multirow{ 6}{*}{Cloud manager}\\ 
\(cldTerm_i\) & Flag for overriding decision of the edge manager on $s_i$ by the cloud for termination& \\ 
\(U_i\) & A set of users to connect to $s_i$ & \\ 
\( l^i\) & The desired application latency objective of $s_i$ & \\
\(Prt_i\) & A list of ports used by $s_i$ & \\ 
\(request_i\) & A list of $[s_i,Pri_i,Prt_i, l_i, U_i]$ to request edge services on an edge node & \\ 
\hline
\( l^n_i\) & Measured average round-trip network latency of $s_i$ & \multirow{ 5}{*}{Monitor}\\ 
\( l^c_i\) & Measured average computing latency of $s_i$ & \\ 
\( l^a_i\) & Computed round-trip application latency of $s_i$; $l^a_i = l^n_i + l^c_i$ & \\ 
\(r_i\) & CPU and memory used by $s_i$ on the edge node; $r_n$ is the CPU and memory used by $s_n$ on the edge node & \\
\(r^r\) & CPU and memory released by terminating servers on the edge node & \\ \hline
\color{black}\(decision\) & \color{black}Flag for 'scaleup' or 'scaledown' containers on an edge node & \color{black}Auto-scaler \\ \hline
\end{tabular}
\label{table:mathnotation}
\end{table*}

In this section, we present provisioning and auto-scaling that are essential to ENORM. 
Provisioning in our framework enables cloud servers to offload workloads on to edge nodes. Auto-scaling takes resource availability on the edge node into account and allocates/de-allocates resources provided to a workload. 
Table~\ref{table:mathnotation} shows the mathematical notation we have used in this paper. 

\subsection{Provisioning}
\label{provisioning}
Procedure~\ref{algo:provisioning} shows the provisioning mechanism of ENORM on an edge node in three stages. They include handshaking, deployment and termination. When an edge manager starts, it is initialised by checking whether the node can support edge services (Line~1). If 
positive,
then a handshaking protocol is initiated (Line~2). {\color{black} Additionally, the cloud manager can terminate its edge server (Lines~4-5).}  

\begin{algorithm}
\SetAlgorithmName{\color{black}Procedure}{}
\DontPrintSemicolon
 \KwData{$service, request_i, cldTerm_i, S, R, r^u, termType_i$}
 \While{\(service == true\)}{
   Handshake($request_i, S, r^u, R$)\;
   Deploy($s_i, U_i,S, R,r^u, termType_i, cldTerm_i$)\;
   \If{\(cldTerm_i == true\)}{
   Terminate$(S, s_i, single, true)$\;
   }
  }
 \caption{Provisioning mechanism}
 \label{algo:provisioning}
\end{algorithm}

\textit{Handshaking}: Procedure~\ref{algo:handshaking} presents the handshaking protocol. The edge node listens for incoming requests from cloud managers. 
If the edge node can provide services to a requesting cloud manager, then the cloud manager is notified and returns an edge server setup request. This request includes the name of the application, priority level, ports required to connect to user devices, users to be served by the edge server and latency objective (provided as $request_i$ by the cloud manager to the edge manager in Procedure~\ref{algo:provisioning}). {\color{black}In ENORM, handshaking is not only identifying an edge node for deploying an application, but also initialising a container on the edge node and setting up appropriate firewalls. 
}

\begin{algorithm}
\SetAlgorithmName{\color{black}Procedure}{}
\DontPrintSemicolon
 \KwData{$request_i, S, r^u, R$}
 \eIf{\(i <= n \And R >= r^u\)}{ \tcp{$S$ is ordered by $Pri$}
   \If{$Prt_i \not\in Prt$}{
   		assign the same no. of ports from $Prt$ to $Prt_i$\;
   		}
    assign an additional port from $Prt$ to $Prt_i$\;
    initialise LXC container on $s_i, Prt_i$\;
    update the firewall of $Prt_i$\; \tcp{\color{black}using iptables command}
    send response to cloud server manager\;
   }{
   reject($request_i$)\;
   }
 \caption{Handshaking mechanism}
 \label{algo:handshaking}
\end{algorithm}

A resource check and a priority check is performed when the setup request is received (basic service takes highest priority and other workloads with a high priority can be executed). If the new request ranked lower than any of the currently executing edge servers or there is not sufficient resource to launch a new container, then the request is rejected (Line~10). The cloud manager may then decide to request services from another edge node. If the request passes the resource and priority checks (Line~1), the edge manager further checks whether the proposed ports are available for supporting communication between the edge server and the user devices (Lines~2-4). An access port is generated (Line~5) for remote access from the cloud server manager. When unique ports are successfully allocated to service the current request, an Operating System (OS) container is launched with a default image offered by the edge manager (Line~6), which will be used for executing the workload offloaded by the cloud server manager. The container is allocated a default amount of resources (for example, one core and 200~MB of RAM). After the container is booted on the edge node, the edge manager configures this container (Line~7), such that the application executed on this container will be visible for connecting user devices as well as for managing by the cloud server. {\color{black} This is done through port forwarding using \textit{iptables} command\footnote{https://linux.die.net/man/8/iptables}}. After configuring the container, a message is sent to the cloud server manager (Line~8). 

\textit{Deployment}: 
Procedure~\ref{algo:deployment} shows deployment on the edge nodes. The cloud server manager deploys a partitioned edge server application and installs software packages on the container required by the server remotely (Lines~1-2). The container is customised for each application server by the cloud server manager, which in turn avoids the wastage of resources if a generic container with comprehensive packages of libraries is booted. Once the application server is launched on the edge node (Line~3) it can start receiving requests from user devices. The cloud server manager redirects the application users covered by the edge node through a configuration file that points to the edge server instead of the cloud server (Line~4). Once users are connected the resources allocated to the container can be scaled (more resource can be allocated if available or de-allocate resources if they are not required; Line~5), which will be considered in the the next section. 

\begin{algorithm}
\SetAlgorithmName{\color{black}Procedure}{}
\DontPrintSemicolon
 \KwData{$s_i, U_i,S, R,r^u$}
 install software packages in $s_i$\;
 deploy partitioned server image in $s_i$\;
 launch $s_i$\;
 redirect $U_i$ to $s_i$\;
 $autoScale(S, s_i, R, U_i,r^u, termType_i, cldTerm_i)$\;
 \caption{Deployment mechanism}
 \label{algo:deployment}
\end{algorithm}

\textit{Termination}: In the fog computing model, it is anticipated that when compared to cloud servers, edge servers will be used for shorter time intervals given the demand and limited resources on edge nodes. Therefore, in addition to auto-scaling, the edge manager will need to decide when an application server on the edge needs to be removed from the edge node as shown in Procedure~\ref{algo:termination}. The edge manager terminates an edge service in the following three cases, which are considered in the auto-scaling mechanism. Firstly, there are no free resources to support the edge service. Secondly, the edge service is not required any more (the edge server has been idle for a time period). Thirdly, the edge service does not improve the QoS of the application (the performance constraints cannot be satisfied by an edge server deployment). The cloud server manager can override the edge server for terminating its edge service, so that when no budget is available for additional edge service usage, the cloud server manager withdraws its server from the edge. 

\begin{algorithm}
\SetAlgorithmName{\color{black}Procedure}{}
\DontPrintSemicolon
 \KwData{\(S, s_i, termType, cldTerm_i\)}
 \If{\( cldTerm_i == true \Or termType == single\)}{
  migrate and redirect $U_i$ to the cloud\;
  stop and destroy LXC container hosting $s_i$\;
  $S = S - \{s_i\}$\;
   }
 \If{$cldTerm_i == false \Or termType == multiple$}{
 	\For{$\forall s_i \in [s_i, s_n]$}{
       migrate and redirect $U_i$ to the cloud\;
  stop and destroy LXC container hosting $s_i$\;
  $S = S - \{s_i\}$\;
  
}
 }
\caption{Termination mechanism}
 \label{algo:termination}
\end{algorithm}

A single or multiple containers can be terminated (based on the value of $termType_i$). 
When an edge application server is terminated, the associated data containing local updates is migrated to the cloud (the local data will be appended to the global data maintained by the cloud server manager; Lines~2 and 8). This is realised through a key-value based data store, Redis\footnote{\url{https://redis.io/}}, which supports data migration between two servers. The user devices affected will have to be redirected back to the cloud server until the next edge node can be provisioned by the cloud server manager.

\subsection{Auto-scaling}
\label{autoscaling}
The importance of scaling resources allocated to an edge application server is in that (i) the edge nodes have limited hardware resources (will need to be primarily used for basic services) and (ii) the application server executing on the node requires more or less resources to the meet the QoS objective {\color{black}(in this paper, we only consider application latency)}. Procedure~\ref{algo:autoscaling} shows the auto-scaling mechanism we have proposed in this paper. This is not a one time method, but happens periodically {\color{black} (in this paper, every 5 minutes)} during the execution of an application server on the edge node. The edge manager, resource allocator and monitor are required to enable auto-scaling. 


\begin{algorithm}
\SetAlgorithmName{\color{black}Procedure}{}
\DontPrintSemicolon
 \KwData{\(S, s_i, R, U_i,r^u, termType_i, cldTerm_i\)}
 \For{$\forall s_i \in S$}{ \tcp{S is ordered by $Pri$}
 $l^n_i = ping(U_i)$\;
 measure $l^c_i$\;
 $l^a_i = l^n_i + l^c_i$\;
 \eIf{$R >= r^u$}{
   \eIf{ $U_i != \emptyset \Or l^n_i < {\color{black}l^i} $}{
   		\eIf{$l^a_i > {\color{black}l^i}$}{
        	$scale(s_i,scaleup, r^u, S, R)$\;
        }{
        	$scale(s_i,scaledown, r^u, S, R)$\;
        }
   }{
   		$terminate(S, s_i, single, false)$\;
   }
   }{ 
   $terminate(S, s_i, multiple, false)$\;
   }
   }
 \caption{Auto-scaling mechanism}
 \label{algo:autoscaling}
\end{algorithm}

A list of application servers executing on the edge node and their priority levels is maintained by the edge manager (Line~1). {\color{black}The priority of an application is set by the cloud manager that owns the application. In this paper, we only consider a static priority.
The priority of the application does not change during execution. If the priority needs to change, then the cloud manager will need to relaunch the application on the edge with a new priority.} 
The application with the highest priority is firstly considered by the auto-scaler.
{\color{black} The network latency is measured by pinging users from the server (Line~2; the reason for choosing pinging is because our demonstrator use case transmits a relatively small amount of data in each HTTP request. However, for other applications, varying data sizes may be more suitable).}
It is then checked if the amount of free resources available on the edge node reported by the resource allocator is larger than the predefined minimum amount of resources required by the edge application server (Line~5). If this resource check fails, then the high priority server along with all other servers that ranked below it will be migrated back to the cloud from where they were offloaded (Line~16).{\color{black} The terminate mechanism used here is the same as presented in Procedure~\ref{algo:termination}.
} This is done to ensure the basic services on the edge nodes retain the highest priority. If the resource check passes, then the auto-scaler further checks if there is a need for the edge application server on the node (whether users are covered by this edge node or whether the latency of the application can be reduced on the edge node; Line~6). The existence of users (whether users are connected to the edge server) and network latency are considered to decide whether the edge server can deliver the desired improvement or if migrating the application server elsewhere to the cloud or to another edge node can be of more benefit (Line~13). 
The computed application latency is compared with the latency objective of the cloud server manager in its service request (Line~7). If the application latency is higher than the service objective, (or the edge server has not been performing as expected), then the container hosting the edge server will be allocated more resources (Line~8). Resources are removed from the container when the latency is less than the objective (Line~10).

\begin{algorithm}
\SetAlgorithmName{\color{black}Procedure}{}
\DontPrintSemicolon
 \KwData{\(r^u, S, decision, R, s_i\)}
 \If{\( decision == scale up\)}{
   Measure \( r_i\)\; \tcp{using LXC command $lxc-cgroup$}
  \eIf{\( R >= r^u\)}{
   \(r_i = r_i + r^u\)\;
   }{ 
   $r^r = 0$\;
   \While{\(R < r^u\)}{
   Measure $r_n$\;
   terminate($s_n, S, single, false)$\;
   $R = R + r_n$\;
   $r^r = r^r + r_n$\;
   $n = n - 1$\;
   \If{\( R >= r^u \Or n==i\)}{
   $ {\bf break} $\;}
   $r_i = r_i + r^r$\;
 }
 }  
   }
   \If{$decision == scale down$ }{
    Measure $r_i$\;
 \(r_i = r_i - r^u\)\;
 }
 \caption{Scaling mechanism}
 \label{algo:scaling}
\end{algorithm}

When an application server is decided to be scaled, the scaling mechanism as shown in Procedure~\ref{algo:scaling} first checks the decision on either "scaleup" or "scaledown" (Line~1 and 20). To scale up, i.e. to allocate more resources, the edge monitor firstly checks if there are additional available resources $R$ on the edge node to support this (Line~3). If there are resources, then one more resource unit \(r^u\) {\color{black}(one core of CPU and 200MB of RAM in this paper)} is added to the container (Line~4). Re-configuring the resource limits of a container is realised through \(cgroup\), i.e. control group, which is a feature of Linux kernel that limits, accounts for and isolates the resource usage (for example, CPU and memory) of a container.
If the available resource is not enough to support scaling up, then the container with lowest priority in the container set $S$ will be terminated to release its resources so that there is more available resources $R$ (Line~9). The containers with lower priorities will be terminated one by one until $R$ is sufficient to support scale up or there are no more containers with lower priorities (Line~13). To scale down, a unit of resources is removed from the current resources that server $s_i$ is allocated (Line~22). At the end of the scaling process, the application edge server \(s_i\) is updated with the new quota of resources (Lines~4, 16 and 22).  
When an application server is either scaled or migrated, the auto-scaler takes the next server from the list to repeat the process above until the resource check fails on an edge node. 

{\color{black} Auto-scaling approaches can introduce instability when the edge node resources are exhausted and if in an auto-scaling round containers with lower priority were scaled down. This can be mitigated if the amount of resources that need to be removed from a container are known beforehand. Our algorithm does not introduce this instability since we do not progressively scale down containers with lower priority. Instead the container with the lowest priority is terminated until there is sufficient resources for a container with the highest priority to scale up.}

\section{An Online Game Use-case}
\label{sec:usecase}
In this paper, we choose a location-aware online game use-case for testing the feasibility of the ENORM framework. Such games are naturally a good fit in the context of fog computing.
For location-aware games the server maintains a global view of connected users. However, this server can be partitioned such that local views specific to a location are managed by edge nodes. The global view will be updated periodically, but less frequently, on the cloud server. Frequent location specific changes are updated locally.

The online-game chosen 
is an open-source implementation of a game similar to Pok\'emon Go\footnote{\url{http://www.pokemongo.com/}}, named iPokeMon\footnote{\url{https://github.com/Kjuly/iPokeMon}}. iPokeMon comprises a client for the iOS platform, which can be used on mobile devices, 
and a server that is hosted on a public cloud. It maps virtual reality on to real world in which users can walk
through streets 
to discover, fight against 
wild virtual creatures, named Pok\'emons, that are geographically distributed. {\color{black} This is latency sensitive since the virtual environment of the user needs to be frequently updated as a user navigates in a location. If the application is serviced from a distant cloud data center, then user experience is affected due to lags in refreshing the environment.}

The Pok\'emon Go server is known to have crashed multiple times during its launch due to extreme user activity\footnote{\url{http://www.forbes.com/sites/davidthier/2016/07/07/pokemon-go-servers-seem-to-be-struggling/\#588a88b64958}}$^{,}$\footnote{\url{https://www.theguardian.com/technology/2016/jul/12/pokemon-go-australian-users-report-server-problems-due-to-high-demand}}. For such a game not only is it essential to replicate servers in different data centers, but given the large number of user connections, an edge layer between the cloud and the user can reduce the distance of communication. 

We designed the iPokeMon game server to be hosted on an Amazon Web Services Elastic Compute Cloud\footnote{\url{https://aws.amazon.com/ec2/}} VM. The server was hosted on a t2.micro\footnote{\url{https://aws.amazon.com/ec2/instance-types/}} VM in the Amazon Dublin data center, which is geographically closest to the authors location in Belfast (an additional server was hosted in N. Virginia for the purpose of comparison). 

\subsection{Implementing the Fog Computing Based Game Using ENORM}
There are three requirements for implementing the above cloud-only iPokeMon as a \textbf{\textit{`fog computing based'}} game. Firstly, the game server needs to be partitioned such that the global view of the system is maintained on the cloud server and the local view on the edge node. {\color{black}In our model, the server is manually partitioned at function level.
Functions related to users and location data are used to create the edge server. 
Only users existing on the cloud server are directed to edge nodes. The functionality to create and authenticate new users resides on the cloud server. The partitioned server residing on the edge node generates location data, such as region code since a user may traverse through new regions that is not associated with the user on the cloud. The edge node then updates this information in the global map, which is used to annotate places in the application client.}

Secondly, iPokeMon needs to be modified so that it can connect to an IP address dynamically. 
{\color{black} The IP address of iPokeMon server is defined in a configuration file maintained by the cloud manager. By default it points to the cloud server, therefore when a user starts to play iPokeMon, the device is connected to the cloud server. Once an edge server is available, the cloud manager updates the configuration file with the IP address of the edge server. When a new user request is created, iPokeMon will switch connection to the edge server according to the configuration file.} 
When the edge server terminates, the user is directed back to the cloud server. 

The third requirement is partitioning the iPokeMon database at run time to support the migration of user-specific data from the cloud server to the edge node. {\color{black} User data is maintained in a {\color{black} Redis} database on the cloud using pre-defined naming standards. Each user has a number of keys in the database which can be filtered. The keys and values related to all users that will connect to a specific edge node are copied to the edge server during deployment. Similarly, when the edge server is terminated, the updated user data on the edge is migrated to the cloud server and merged with the database containing the global view.}

For executing the fog computing based iPokeMon game, data will be sent from a user device to the game server through a traffic routing node, such as a mobile base station. 
We used an ODROID-XU board\footnote{\url{http://www.hardkernel.com/}} to represent the computing resources of a small cell base station. The board has one ARM Big.LITTLE architecture Exynos 5 Octa processor and 2~GB of DRAM memory. The processor runs Ubuntu 14.04 LTS. This Odroid board served as the edge node, which was located in the Computer Science Building of the Queen's University Belfast in Northern Ireland. 

\begin{figure}
  \centering
  \includegraphics[width=0.49\textwidth]{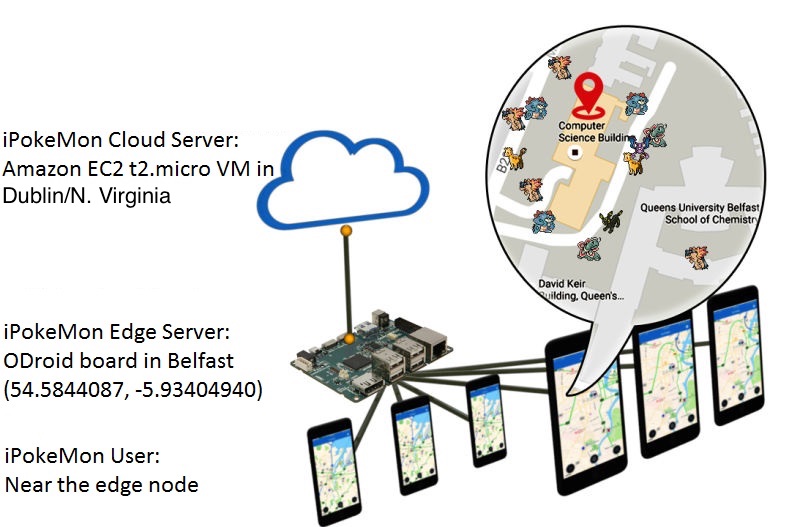}
  \caption{The implementation of fog computing based iPokeMon game using the ENORM framework. The cloud server is on Amazon EC2 and the edge server is on an Odroid board that connects user devices.}
  \label{fig:edgeBasediPokeMon}
\end{figure} 

Figure \ref{fig:edgeBasediPokeMon} shows our implementation of the fog computing based iPokeMon game using ENORM. 
The user creation and verification requests are made to the cloud server, after which the cloud manager makes a request for computing services to a potential edge node. Following this handshaking described in Section~\ref{provisioning} is established. If this request is accepted by the edge node, then the edge manager initialises a container for the iPokeMon edge server. The cloud manager deploys the iPokeMon edge server and clones the data (to the edge database) of the users that will be connected to the edge node. 
User data rapidly changes when the game is played. For example, the GPS coordinates of the player and the Pok\'emons. The local view on the edge server is updated by frequent update requests that are sent to the edge server. 
When the edge server has to be terminated, 
as considered in Section~\ref{provisioning}, 
then the edge database is merged with the global database located on the cloud. The user is redirected back to the cloud server for continuing the session. 
If a new edge node is available, then the above process is repeated.

\section{Experimental Studies}
\label{sec:evaluation}
In this section, we evaluate our fog computing based implementation of the iPokeMon game. 
The partitioned game servers on the edge node 
were stress tested using Apache JMeter\footnote{\url{http://jmeter.apache.org/}}. One session of a connection (the user is playing the iPokeMon game) between the user device and the edge server hosted on a container is recorded for 20 minutes. During this time the number and types of requests and the parameters sent through the requests are recorded. Subsequently, JMeter stress tests single and multiple edge servers by creating virtual users and sending requests to the edge server(s) from the virtual users in the experiments. 

{\color{black}The activity of `$N$' virtual users
is defined by considering two types of user behaviour. Firstly, aggressive user behaviour, in which the activity of each user is considered by randomly selecting only data intensive requests with no pauses between these requests for a 5-minute period from the pre-recorded 20 minutes. {\color{black}For example, when a user is continuously playing by rapidly tapping the game screen with few breaks. All moves and taps made by the user will need to be transmitted from the phone to the game server.} This is highly unrealistic, but is employed to represent a bandwidth-hungry task. Secondly, mixed user behaviour, in which the activity of each user is considered by equally selecting both data intensive requests (with pauses) and regular requests for a 5-minute period from the pre-recorded 20 minutes. {\color{black}For example, when a user plays iPokeMon strategically, by taking time for a next move before tapping the game screen. Some moves may require a very small amount of data to be transmitted while others may require larger amounts of data.} This ensures different behaviours of the virtual users and captures a real world setting in which there are a combination of aggressive and passive users.} 

For comparing the cloud-only and fog computing based implementation using ENORM we have stress tested the cloud server.
Each experiment shown in this paper is based on the stress test for 5 minutes and represents an average of 5 executions. {\color{black} To simulate the effect of a dynamic basic service (utilisation of resources on the edge node changes) on the free resources $R$, an open dataset\footnote{http://crawdad.org/ctu/personal/20120315/} containing 142 days of mobile phone records was used to extract hourly patterns. We considered bursty patterns of basic services representative of variable resources utilisation on edge nodes. }

Our experimental evaluation demonstrates the following:
(i) the overheads in provisioning and auto-scaling using ENORM for the iPokeMon game, (ii) the improvement in the QoS when using ENORM measured by (a) user perceived latency of communication between the device and the edge node, (b) the data transfers between the user device and the cloud server, and (c) the communication frequency between the user device and the cloud server.   

\subsection{Overheads}
We initially determined the overhead in setting up edge server using ENORM. This provides us with an insight into the expense incurred for using edge server as well as determining the impact of bringing about changes, such as auto-scaling, to the edge server, in terms of time. Here we explore the overheads for provisioning and auto-scaling. 


\begin{figure*}
\begin{center}
	\subfloat[For a single user on each container]
	{\label{fig:figure1a}
	\includegraphics[width=0.49\textwidth]
	{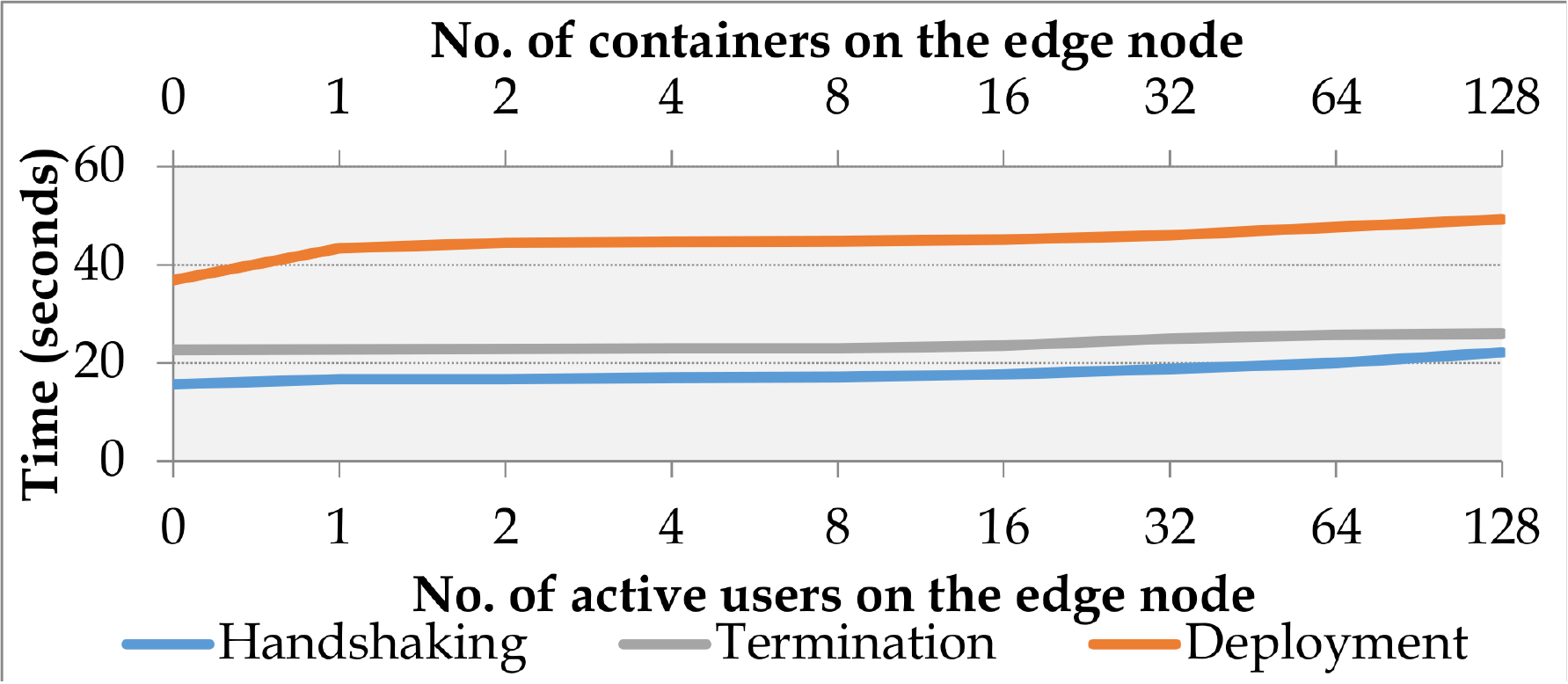}}
\hfill
	\subfloat[For multiple users on each container]
	{\label{fig:figure1b}
	\includegraphics[width=0.49\textwidth]
	{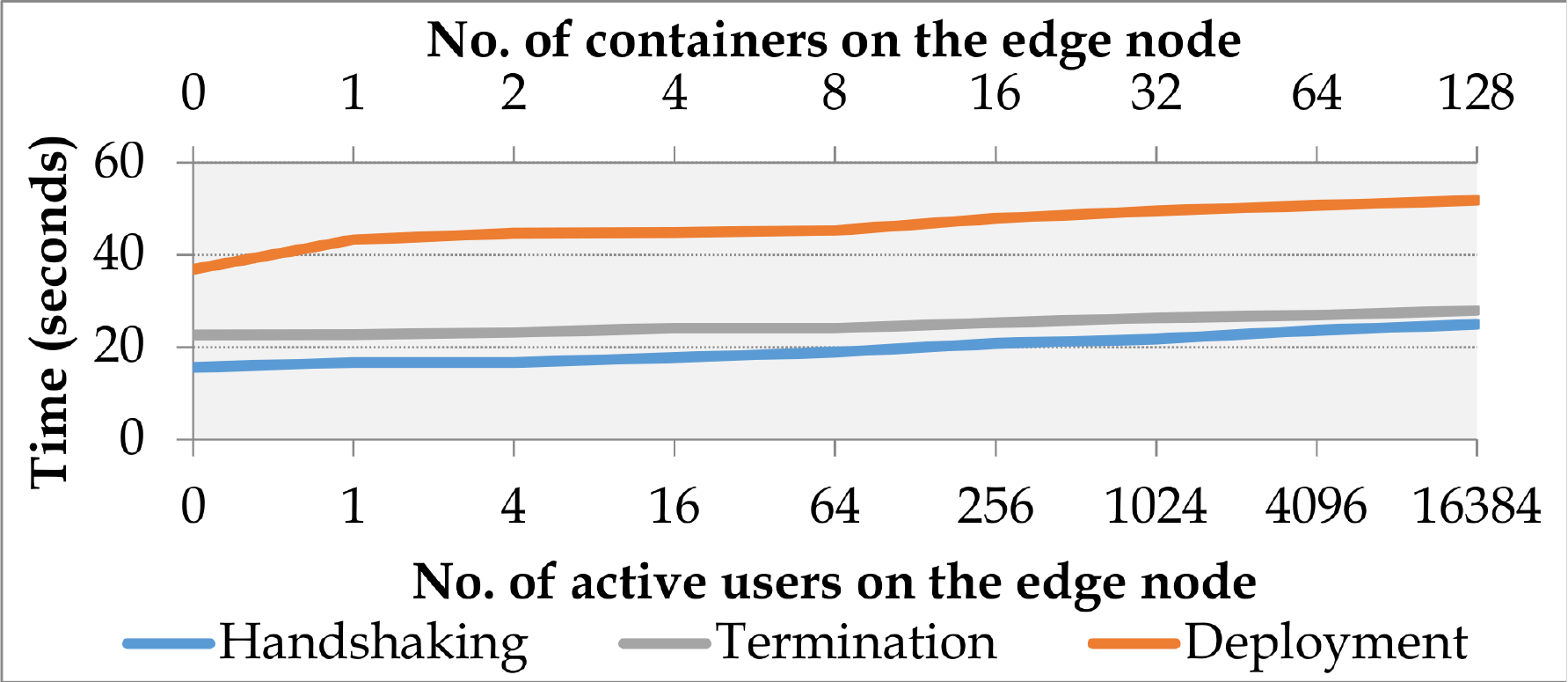}}
\end{center}

\caption{Overheads in provisioning edge servers using ENORM on an Amazon EC2 server located in the Dublin data center.}
\label{fig:figure1}
\end{figure*}

\begin{figure*}
\begin{center}
	\subfloat[For a single user on each container]
	{\label{fig:figure2a}
	\includegraphics[width=0.49\textwidth]
	{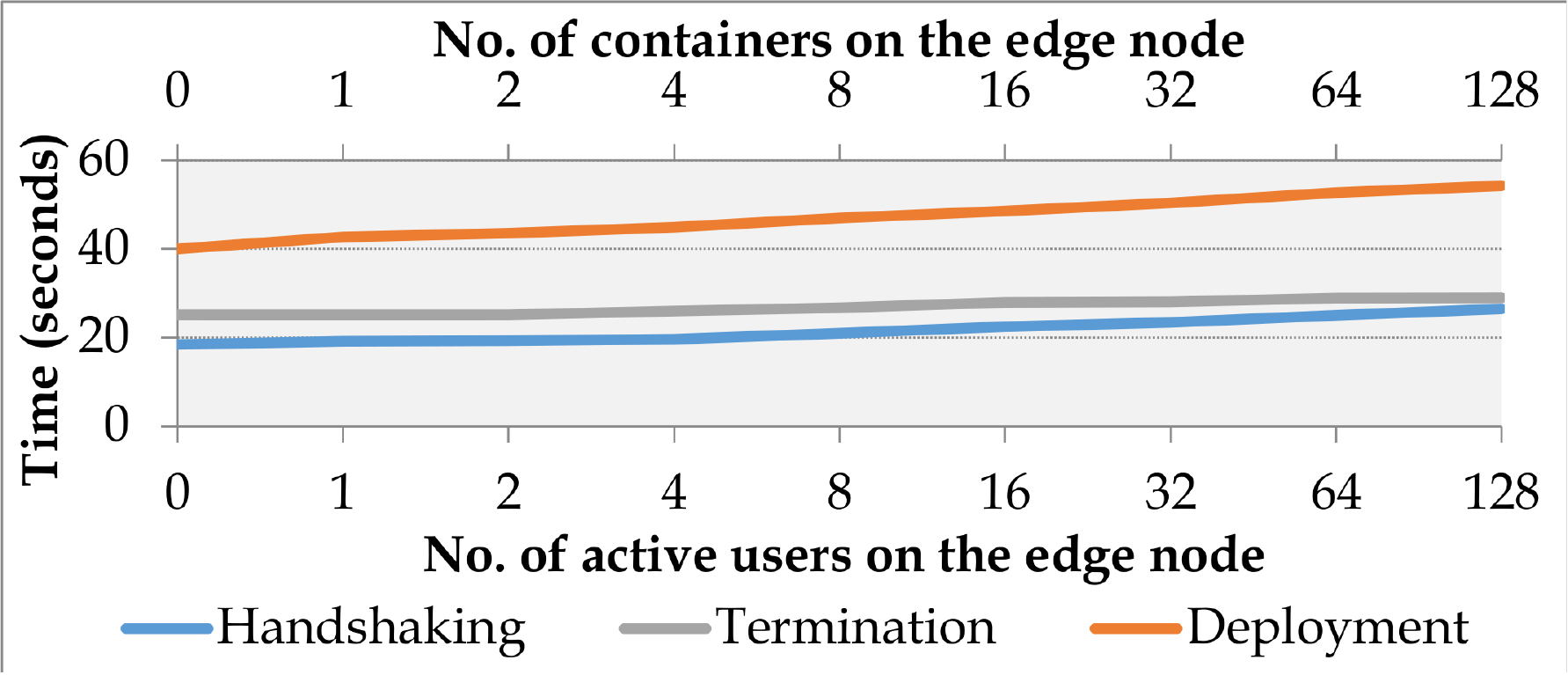}}
\hfill
	\subfloat[For multiple users on each container]
	{\label{fig:figure2b}
	\includegraphics[width=0.49\textwidth]
	{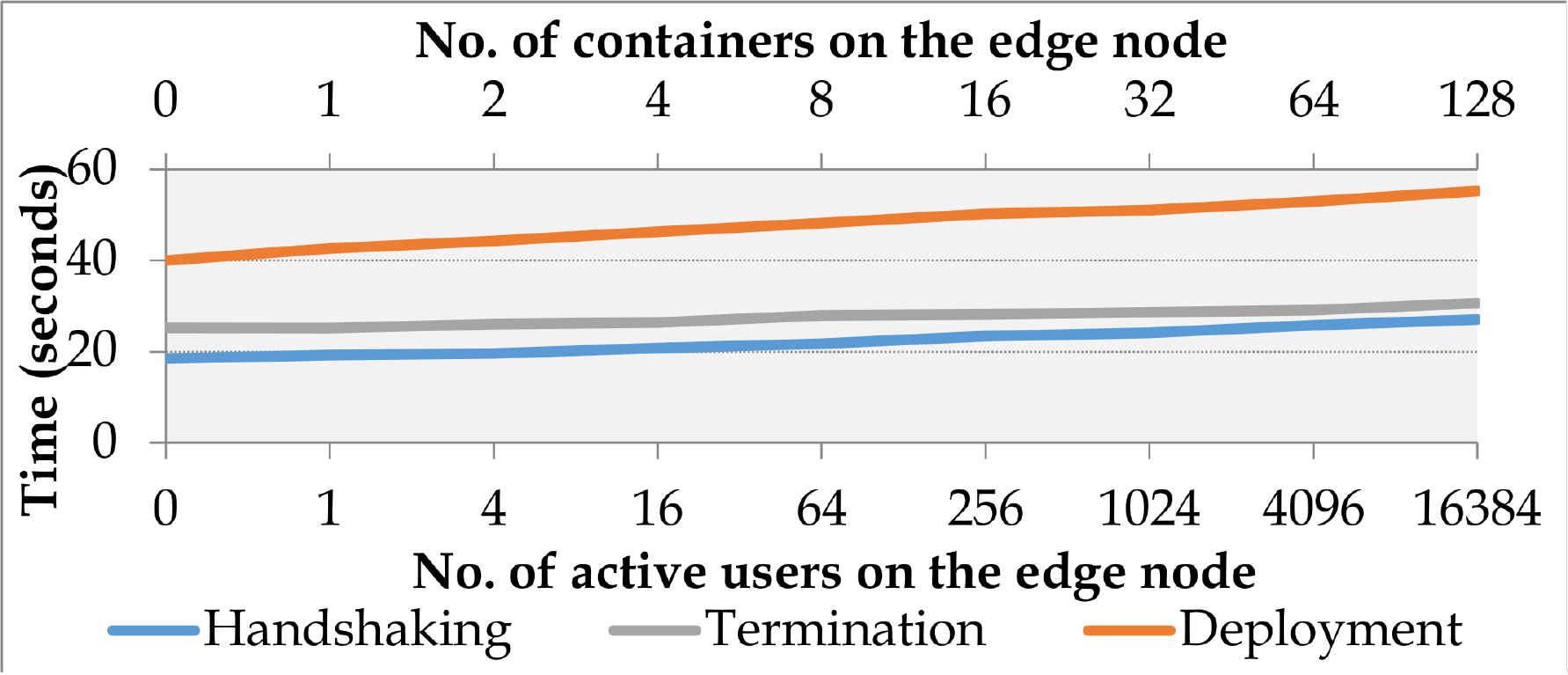}}
\end{center}

\caption{Overheads in provisioning edge servers using ENORM on an Amazon EC2 server located in the N. Virginia data center.}
\label{fig:figure2}
\end{figure*}

\textit{Provisioning:} Figure~\ref{fig:figure1} and Figure~\ref{fig:figure2} show the overhead in provisioning iPokeMon edge servers using ENORM on a cloud server located in the Dublin and N. Virginia data centers respectively. As presented in Section~\ref{provisioning}, there are three stages, namely handshaking during which the edge node is made available for hosting the partitioned server, deployment during which the application server is deployed on the edge node container, and termination during which control from the edge node is given back to the cloud server. We observed the overhead for each stage. 

The evaluation takes two scenarios into account. Firstly, when only one user is connected to each server (Figure~\ref{fig:figure1a} and Figure~\ref{fig:figure2a}) and when multiple users are connected to each server (Figure~\ref{fig:figure1b} and Figure~\ref{fig:figure2b}). 
Given $n = 0, 1, 2, \cdots, 128$ containers executing on the edge node, we obtained the handshaking overhead when the cloud server needs to establish connection with a new container on the edge node. {\color{black}In this paper, we define handshaking overhead as the time taken for identifying an edge node for deploying an application, initialising a container on the edge node and setting up appropriate firewalls.
Similarly, the deployment overhead when a new application server is deployed on the edge node and the termination overhead when a container is terminated on the edge node are obtained.} The general trend is that there is a slight increase in the overheads when there is more stress on the edge node. This is expected, but it is worthwhile to note that there is less than a 3\% increase in the overheads when multiple users connect to the edge servers in comparison to when single users are connected (compare Figure~\ref{fig:figure1a} and Figure~\ref{fig:figure1b}). This overhead is because the users' information needs to be transferred from the cloud to the edge during deployment and from the edge to the cloud during termination, which is a single state information, rather than the intermediate states which are maintained on the edge node. This highlights that ENORM can handle a large number of user connections and edge servers for the iPokeMon use-case. 

The overheads when using the cloud server in the N. Virginia data center is higher than in Dublin because the handshaking requires more time to communicate given the geographic distance (similarly for termination and deployment). {\color{black}The deployment overhead is under 1 minute in all cases. Until the edge node server is deployed and fully running, the users are still connected to the cloud server and user requests are serviced by the cloud. A user is redirected to the edge node only after the server is deployed. }

The deployment overhead is higher than the handshaking and termination overheads. This is because in ENORM, each time a new server is instantiated using a container on the edge node, {\color{black} new packages relevant to the application are installed in the container to host the partitioned server. A custom image of the container hosting the partitioned server is not created by the cloud manager, instead a lean container with a basic image suited for an edge node processor architecture is used. For example, an Alpine Linux image suited for the ARM architecture is used on the Odroid board, but a different image would be required for an Intel architecture edge node. Any additional packages specific to an application needs to be installed by the cloud manager during deployment. Given that edge nodes are less likely to be uniform environments with homogeneous processors, ENORM assumes heterogeneity of edge nodes and installs containers suited to the edge node processor architecture. A pre-built image specific to an application may be used by the cloud manager, but requires knowledge of all processor architectures that will be employed at the edge.}


\begin{figure}
	\begin{center}
		\includegraphics[width=0.49\textwidth]{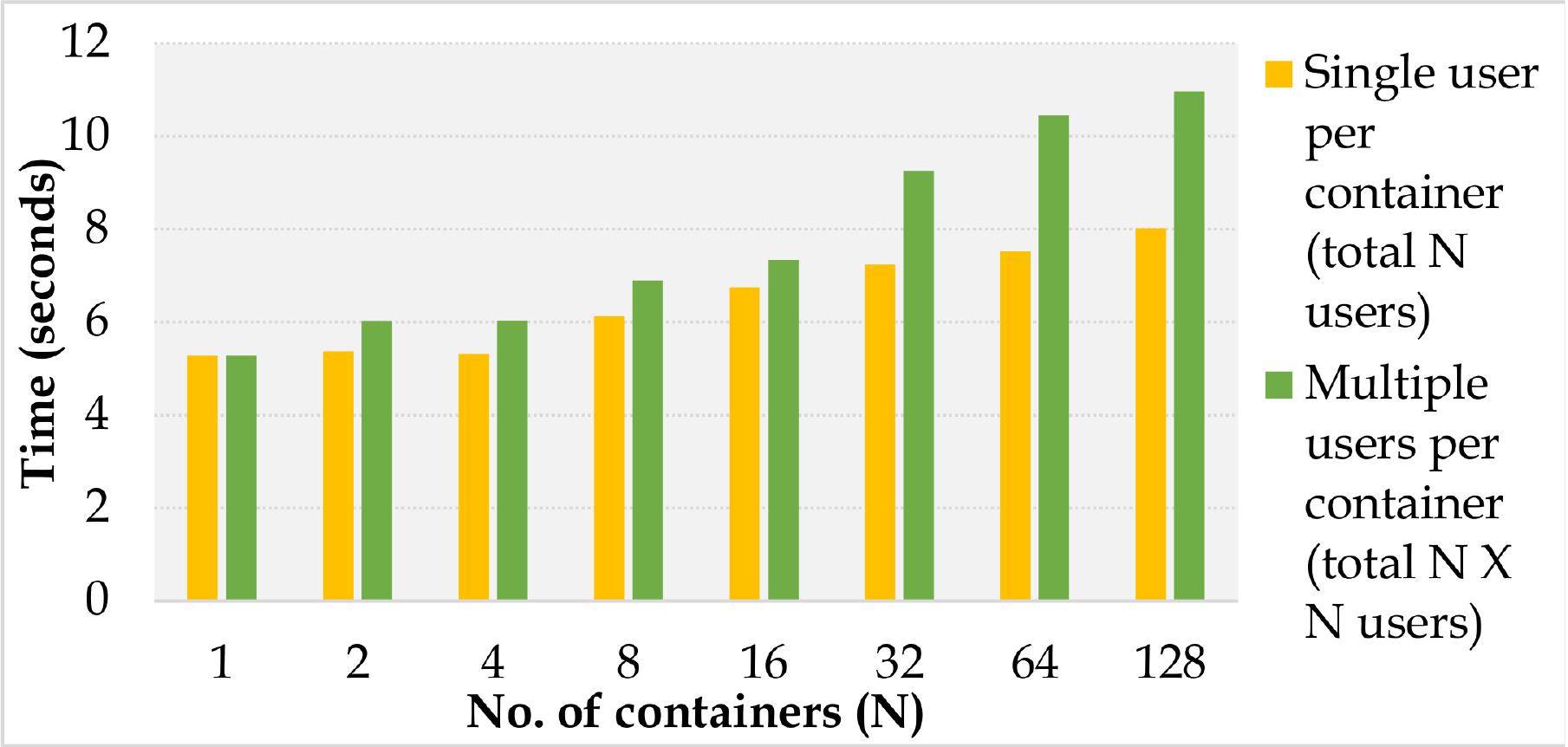}
	\end{center}
    \caption{Total overhead for auto-scaling $N$ containers on the edge node.}
    \label{fig:figure3}
\end{figure}

\begin{figure*}
\begin{center}
	\subfloat[3G network]
	{\label{fig:figure8a}
	\includegraphics[width=0.325\textwidth]
	{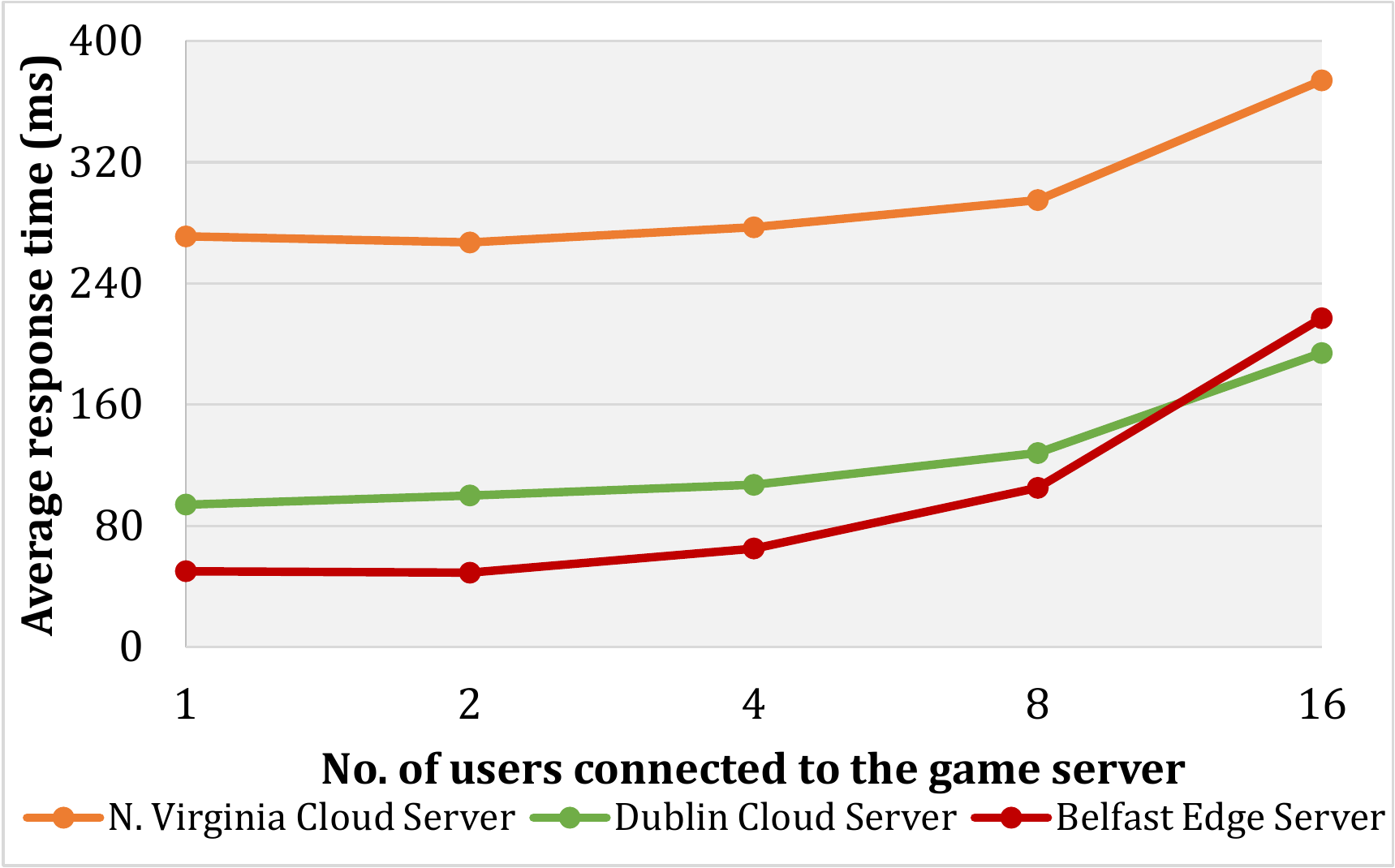}}
\hfill
	\subfloat[4G network]
	{\label{fig:figure8b}
	\includegraphics[width=0.325\textwidth]
	{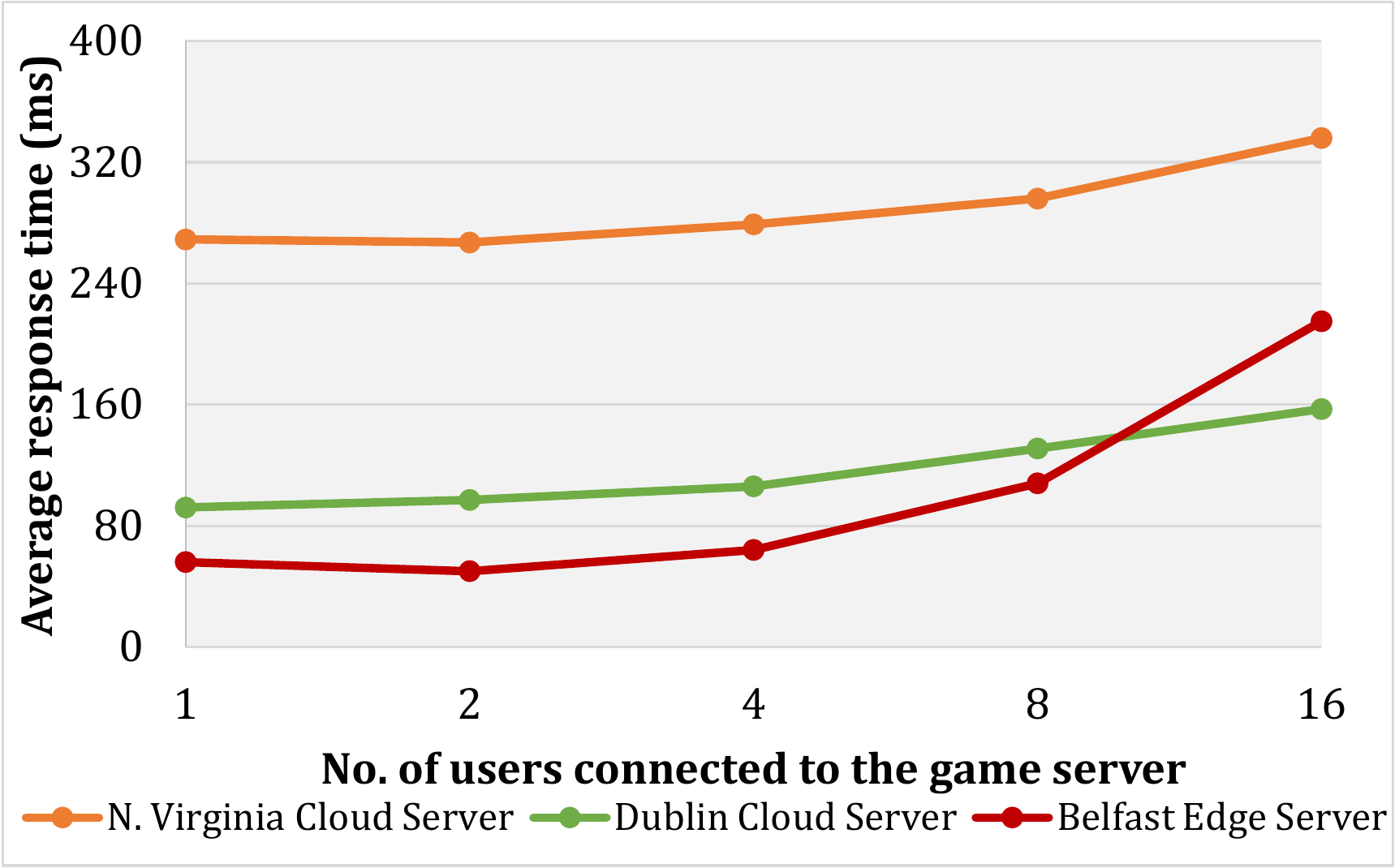}}
\hfill
	\subfloat[Wi-Fi network]
	{\label{fig:figure8c}
	\includegraphics[width=0.325\textwidth]
	{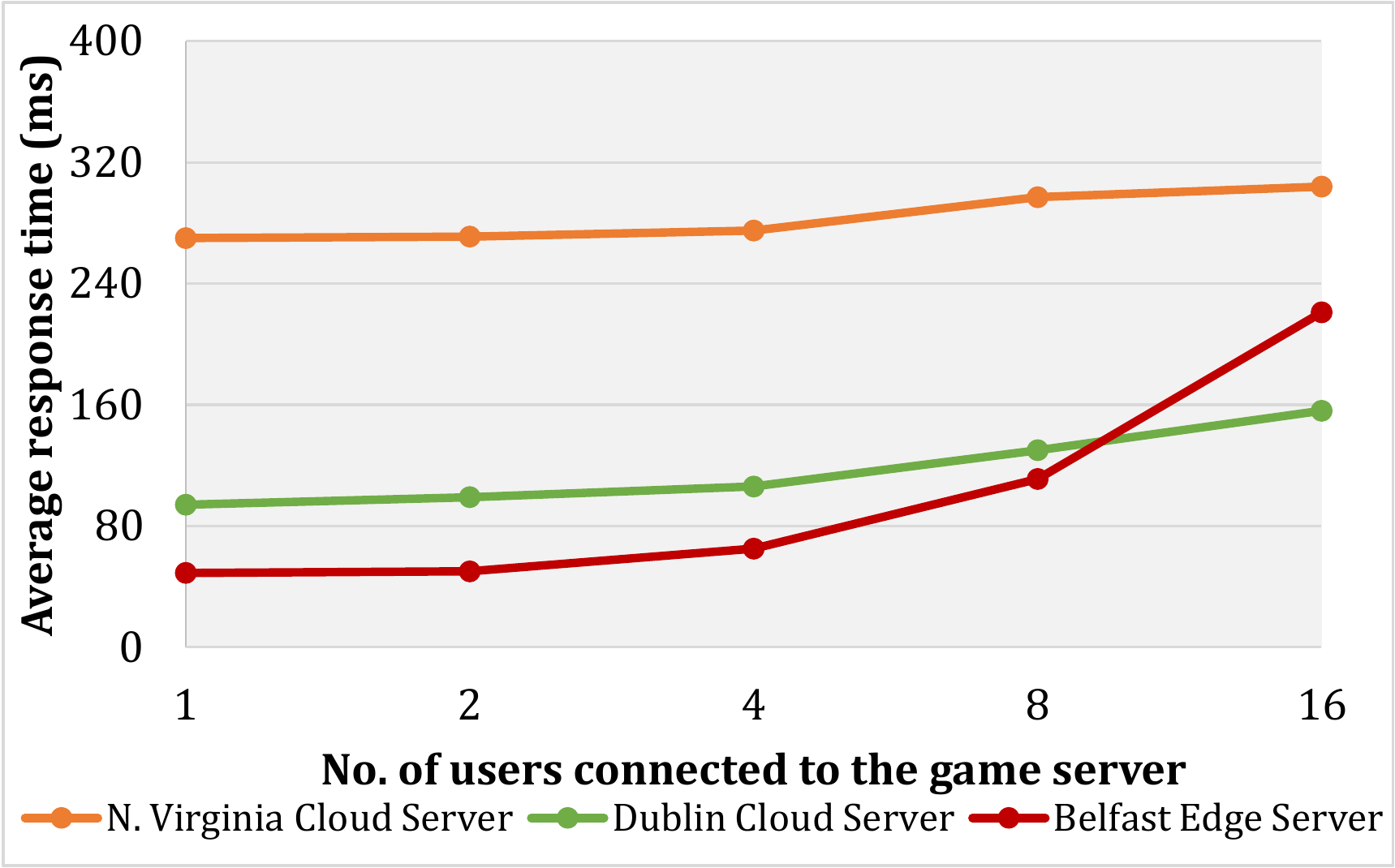}}
\end{center}
\caption{\color{black} Comparing application latency in the cloud-only and fog computing based iPokeMon game with aggressive user behaviour.}
\label{fig:figure8}
\end{figure*}

\begin{figure*}
\begin{center}
	\subfloat[Single server, multiple users]
	{\label{fig:figure5a}
	\includegraphics[width=0.325\textwidth]
	{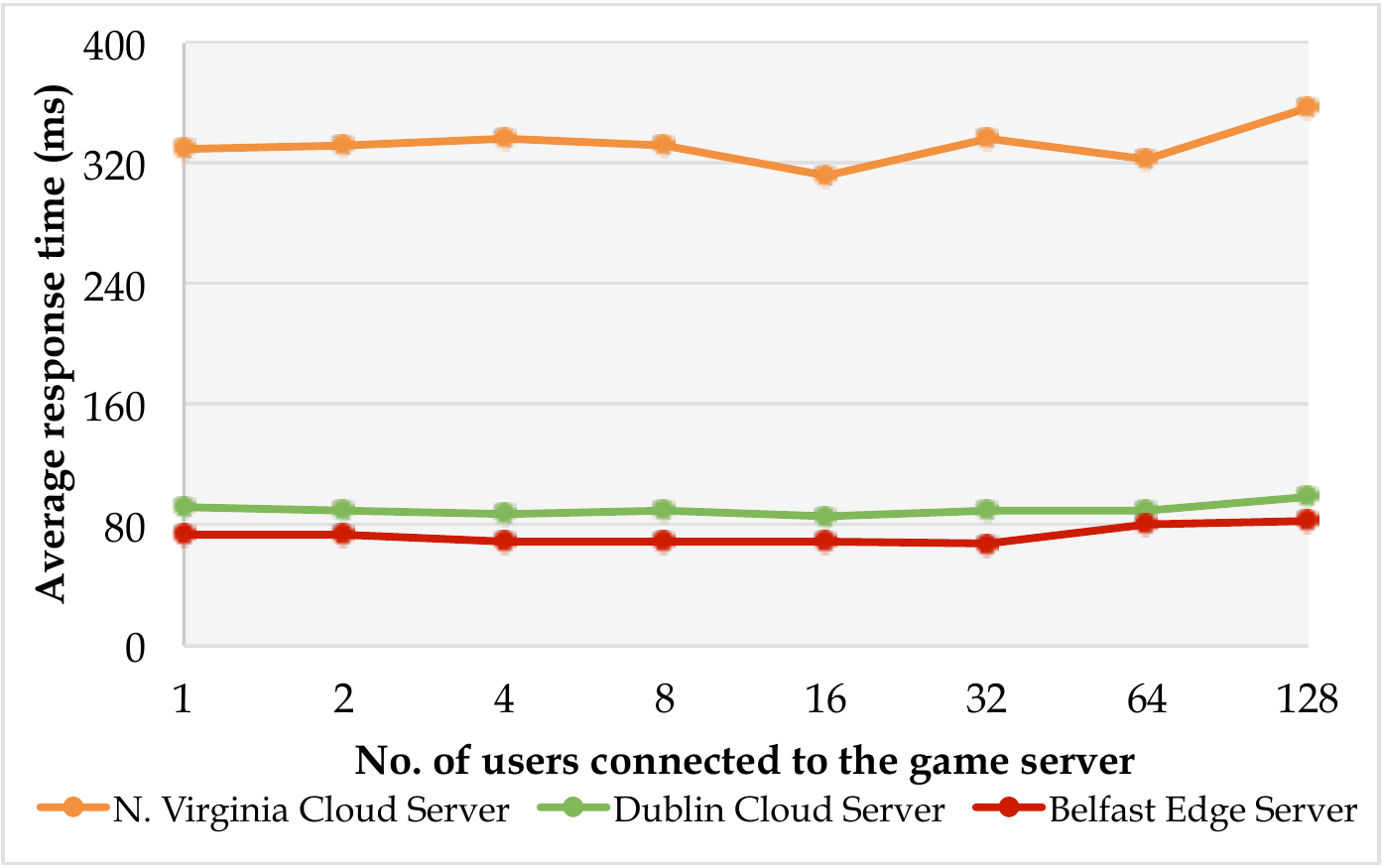}}
\hfill
	\subfloat[Multiple servers, single user]
	{\label{fig:figure5b}
	\includegraphics[width=0.325\textwidth]
	{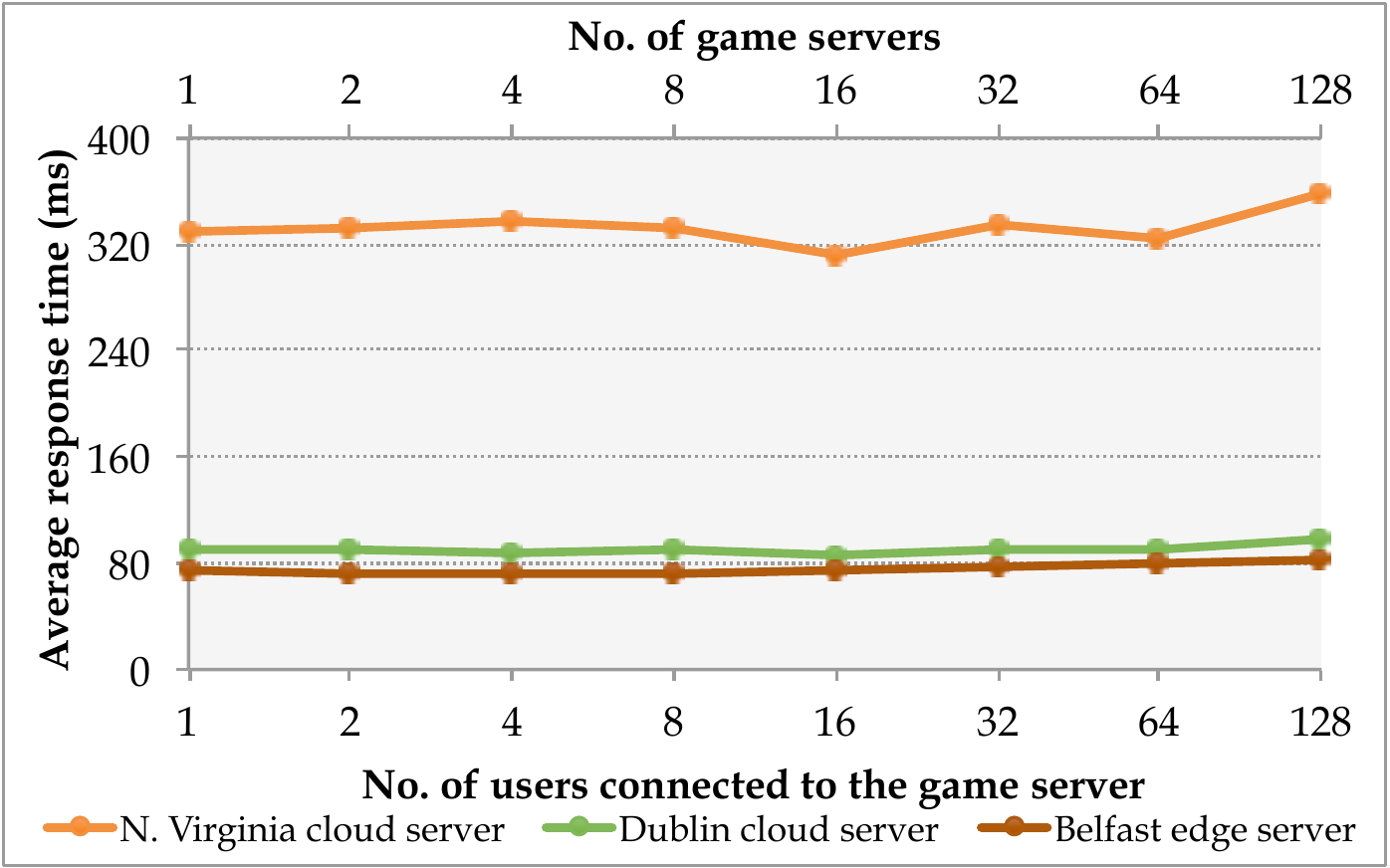}}
\hfill
	\subfloat[Multiple servers, multiple users]
	{\label{fig:figure5c}
	\includegraphics[width=0.325\textwidth]
	{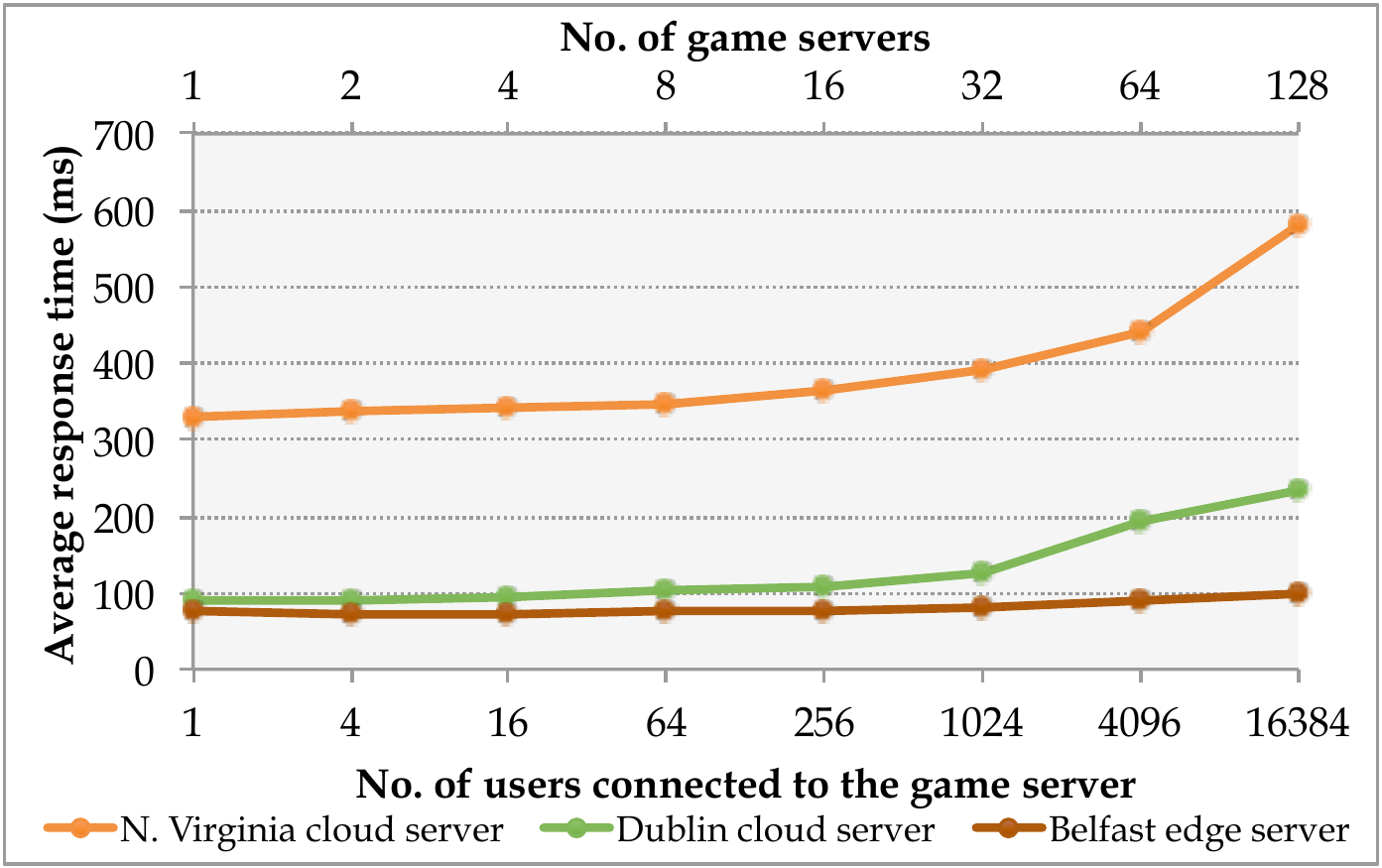}}
\end{center}
\caption{Comparing application latency in the cloud-only and fog computing based iPokeMon game with mixed user behaviour.}
\label{fig:figure5}
\end{figure*}

\textit{Auto-scaling}: For evaluating the overhead associated with auto-scaling we considered the following two scenarios. Firstly, a single user per container (only one user is connected to an edge server, up to a maximum of 128 users on the edge node). Secondly, multiple users per container (given N edge servers, N users are connected per server, up to a maximum of 16,384 users on the edge node). In both scenarios, the experiment started with N containers. For auto-scaling there are three considerations: (i) a container needs to scale up given that it has a high priority and needs to improve the service it offers, (ii) a container needs to scale down, but its application latency will not be affected, and (iii) a container needs to be terminated on the edge node given its lower priority. Using the ENORM framework Procedure~\ref{algo:autoscaling} is executed by the auto-scaler. Consequently, containers with a higher priority scale up, a number of containers scale down since removing resources do not affect the service it offers and containers are terminated. 

Figure~\ref{fig:figure3} shows the total time taken to execute the auto-scaling mechanism over $N$ containers in the system. When the edge server has only one container there is an overhead of 5.3 seconds. However, when more containers are executed on the system, there is additional overhead {\color{black}(compare bars of the same color)} in monitoring all the containers.
Increasing the number of users in a container also affects the overhead {\color{black}(compare yellow and green bars)} because more users have to be monitored. The auto-scaling overhead impacts decision-making on how frequently auto-scaling can be performed on an edge node for a given workload. In this case, for 128 containers there is an overhead of up to 11 seconds. It would not be worthwhile to auto-scale, for example every 30 seconds or 1 minute, which will render the system unstable, but may be appropriate for larger time intervals to update changing priorities of applications.  

\begin{figure*}
\begin{center}
	\subfloat[Mixed user behaviour]
	{\label{fig:figure6a}
	\includegraphics[width=0.49\textwidth]
	{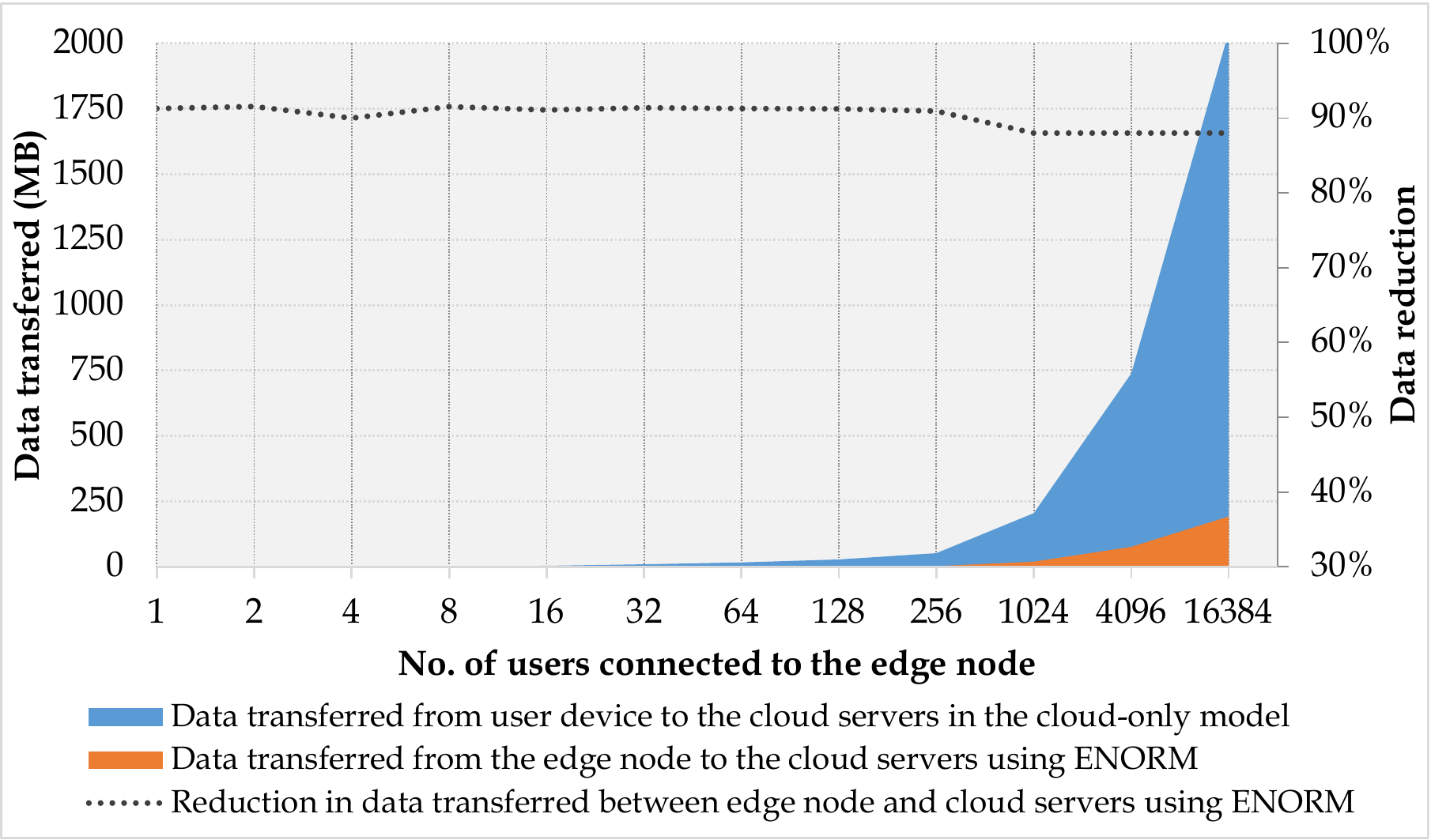}}
\hfill
	\subfloat[\color{black} Aggressive user behaviour]
	{\label{fig:figure6b}
	\includegraphics[width=0.49\textwidth]
	{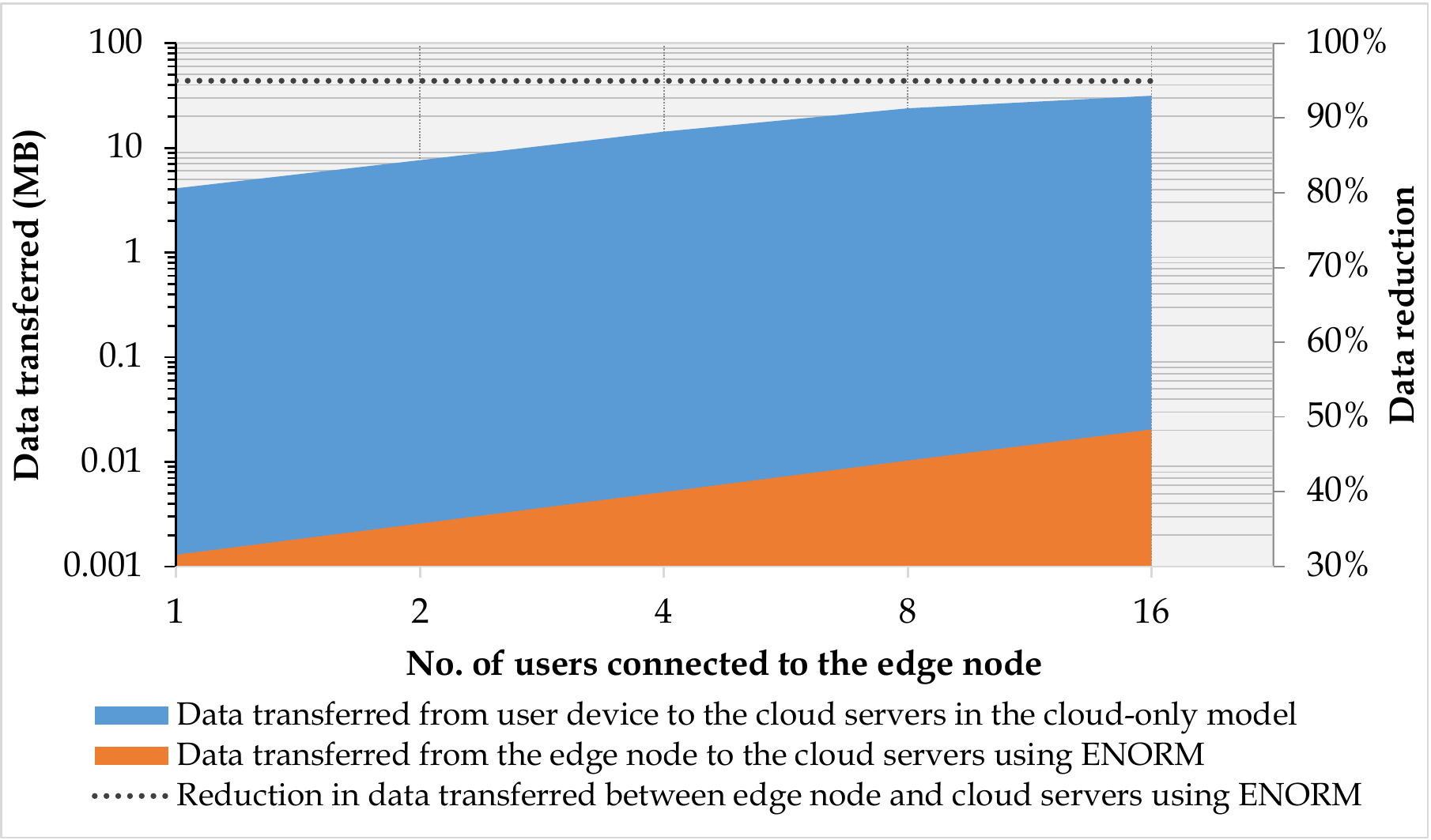}}
\hfill
\end{center}
\caption{Comparing the amount of data transferred in the cloud-only and fog computing based iPokeMon game.}
\label{fig:figure6}
\end{figure*}




\subsection{Service improvement using ENORM}
The benefit of using ENORM to achieve the fog computing based iPokeMon is highlighted in terms of reduced application latency, reduced data transfer between a user device and the cloud server and reduced frequency of communication between a user device and the cloud server. 

\textit{Application latency}:
{\color{black} To evaluate application latency, both aggressive and mixed user behaviours are considered. 
Aggressive user behaviour demonstrates a bandwidth-hungry task. 
Figure~\ref{fig:figure8} shows the application latency for iPokeMon with aggressive user behaviour when using the ENORM framework in a single server with multiple users scenario. Using JMeter we considered three network configurations: (i) 3G network; refer Figure~\ref{fig:figure8a}, (ii) 4G network; refer Figure~\ref{fig:figure8b}, and (iii) Wi-Fi network; refer Figure~\ref{fig:figure8c}. All three network configurations seem to have little influence on the application latency. {\color{black}This is because we did not exhaust the bandwidth in every HTTP request. The results may be different for other use cases that will exhaust the bandwidth.} It is found that the edge server performs better than the cloud servers until there are more than 8 concurrent users. The application overhead increases beyond 8 users. Therefore, in this context for bandwidth-hungry tasks for iPokeMon with more than 8 aggressive users and a total of 73.3 requests per second it is not useful to offload on to an edge server.}

{\color{black}To demonstrate a real world setting, mixed user behaviour is considered where the collection of user requests includes the transfer of small and large data and accounts for pauses between requests. Since the network configuration had little influence on application latency, we present all further results with 4G network configuration. }

Figure~\ref{fig:figure5} shows the reduction in application latency for iPokeMon {\color{black} with mixed user behaviour} when employing ENORM. Three cases are considered: (i) using a single server with multiple users; refer Figure~\ref{fig:figure5a}, (ii) using multiple servers with a single user; refer Figure~\ref{fig:figure5b}, and (iii) using multiple servers with multiple users; refer Figure~\ref{fig:figure5c}. For comparison we employed two cloud servers, the first is located in Dublin, which is closest to the location of our edge node and the second is located in N. Virginia. In all three cases, it is immediately inferred that despite the increasing number of servers or users there is a reduction in the application latency. In the cases of a single server (multiple users) and multiple servers (single user), approximately 20\% reduction is noted in the application latency when using the fog computing model and comparing against the Dublin server. The latency for the application further decreases by nearly 80\% if the N. Virginia cloud server is used (this is because N. Virginia is geographically further from Dublin). 

Figure~\ref{fig:figure5c} shows a more realistic case when there are multiple application servers serving multiple users. In this case, it is observed that for 16,384 users a single request is serviced in 600~ms by the N. Virginia cloud server. On the other hand, the edge server can furnish this request in less than 100~ms. This is a 83\% improvement. Similarly, the application latency is reduced for the Dublin server.

\textit{Data transfer}:
Figure~\ref{fig:figure6} shows the reduction in the amount of data transferred between an edge node and the cloud server in the iPokeMon use-case {\color{black}for mixed user behaviour (Figure~\ref{fig:figure6a}) and aggressive user behaviour (Figure~\ref{fig:figure6b})}. It is observed that using ENORM the partitioned game server can process the data generated by user devices. On average the data transferred between the edge node and the cloud server is reduced by over 88\% {\color{black} for mixed user behaviour} and up to 
{\color{black} 95\% is achieved for aggressive user behaviour (although the edge server is not scalable)} in our use-case. This is encouraging in the context of a large volume of data generating devices that will be connected to the Internet.
ENORM can facilitate computing closer to the device at the edge node layer such that very little traffic is generated beyond the first hop of the network. For example, when 16,384 users connect to the iPokeMon server from the same location for 5 minutes, 2,000~MB of data is generated. In the cloud-only model all data is sent to the cloud server, but using ENORM only 190~MB of data is sent to the cloud beyond the edge node. {\color{black} This corresponds to approximately 12 KB data per user (190 MB / 16,384 users) transferred between the edge node and cloud during the redirection from the cloud server to the edge server, which includes a configuration file containing the dynamic IP address of the edge node and the data of the user in the local view.}

\begin{figure*}
\begin{center}
	\subfloat[Mixed user behaviour]
	{\label{fig:figure7a}
	\includegraphics[width=0.49\textwidth]
	{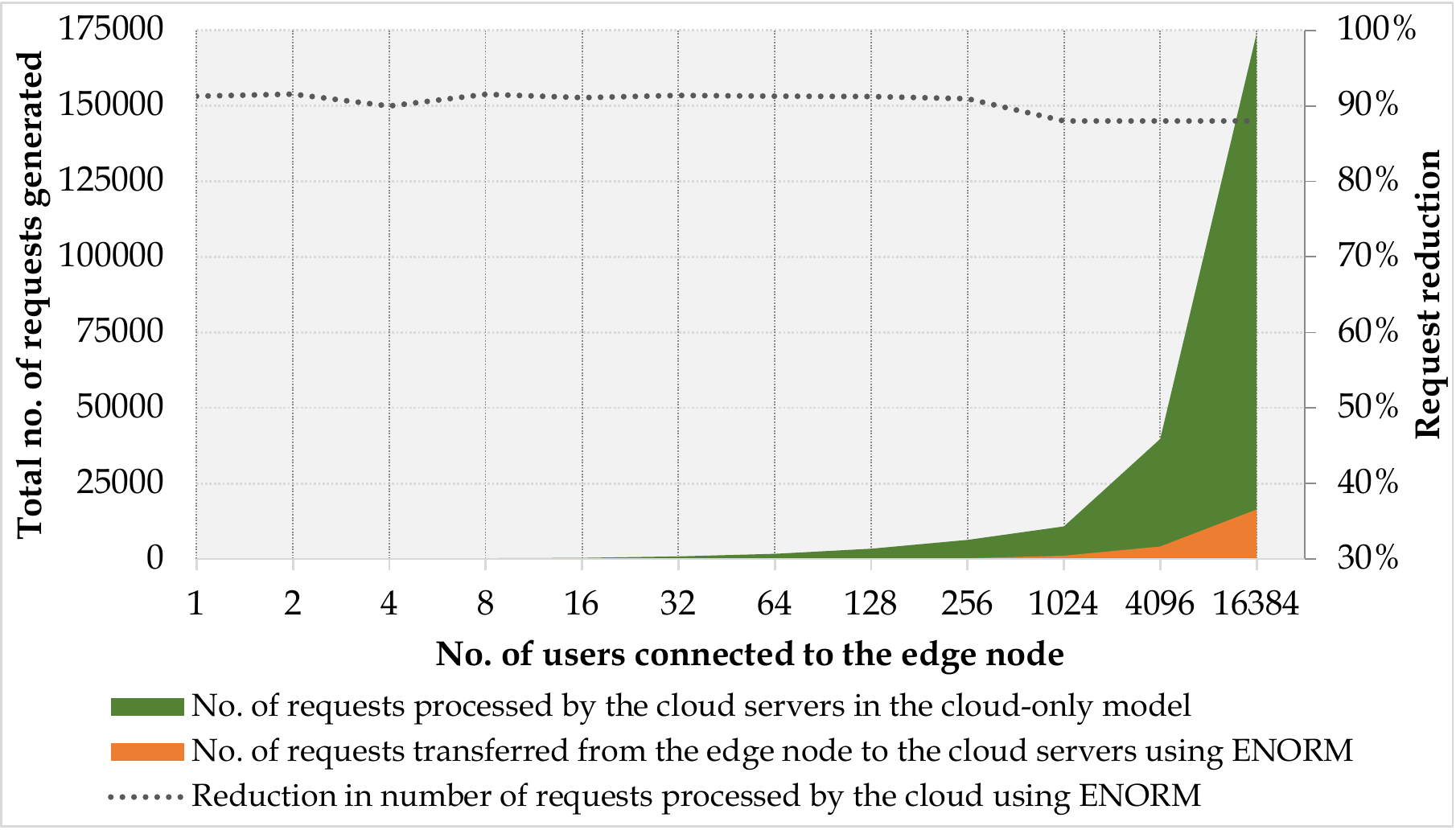}}
\hfill
	\subfloat[\color{black} Aggressive user behaviour]
	{\label{fig:figure7b}
	\includegraphics[width=0.49\textwidth]
	{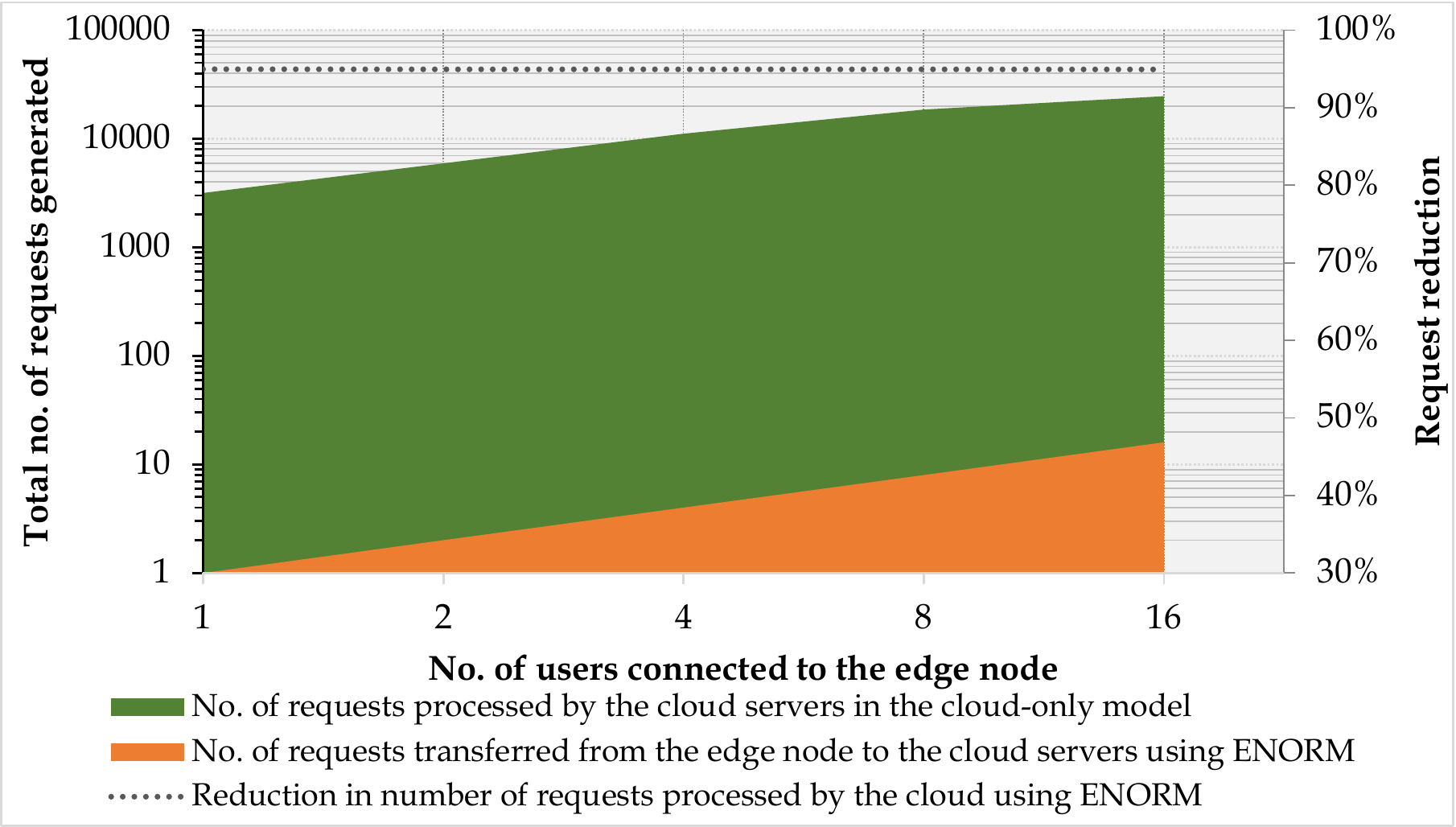}}
\hfill
\end{center}
\caption{Comparing the communication frequency in the cloud-only and fog computing based iPokeMon game.}
    \label{fig:figure7}
\end{figure*}

\textit{Communication Frequency}:
Figure~\ref{fig:figure7} shows the reduced frequency of communication between an edge node and the cloud server in the iPokeMon use-case {\color{black}for mixed user behaviour (Figure~\ref{fig:figure7a}) and aggressive user behaviour (Figure~\ref{fig:figure7b})}. This is similar to data transfer shown in Figure~\ref{fig:figure6}. The number of requests generated in a 5-minute interval for varying number of active users is reduced by 88\%-95\% using ENORM when compared to the cloud-only model. For example, when 16,384 active users connect to the iPokeMon server from the same location for 5 minutes, nearly 175,000 requests are sent to the cloud server in the cloud-only model. Using ENORM, the edge node can service over 90\% requests and only forwards less than 10\% requests to the cloud server in the fog computing model. 
Again the benefit of ENORM in the fog computing model is obvious in reducing the number of user requests that needs to be serviced by the cloud. Only between 8\%-12\% of requests need to be forwarded from the edge to the cloud server.  

\subsection{Summary}
We summarise the experimental evaluation as follows. Firstly, there are provisioning (handshaking, deployment and termination) and auto-scaling overheads when using ENORM. 
It is noted for the use-case that there is only a small increase in the overheads; the overhead for handshaking (9~seconds), deployment (9~seconds), termination (5~seconds) and auto-scaling (6~seconds) when moving from one user to 16,384 users in the edge node. 

Secondly, ENORM improves the QoS of iPokeMon. This is observed by reducing the application latency up to a maximum of 95\% in the best case and reducing data transfer and communication frequency between the edge node and the cloud both by up to 95\%.

\section{Related Work}
\label{sec:relatedwork}


There are numerous challenges in managing resources in a distributed computing environment. Mechanisms to address these challenges have been developed and explored for different environments, such as grids~\cite{gridmanagement-2},
clusters~\cite{
clustermanagement-2}, and more recently on the cloud~\cite{
cloudmanagement-2}. 

The key challenges addressed in relation to managing cloud resources, for example, include (i) scheduling for efficiently mapping workloads on to computing resource~\cite{
cloudscheduling-3}, (ii) benchmarking for selecting computing resources most suitable for a workload at a given time~\cite{
benchmarking-4}, (iii) monitoring for tracking performance of cloud resources against service agreements and user objectives~\cite{cloudmonitoring-1},
and 
(iv) resource scaling for adding or removing cloud resources to meet the computational and storage demands of a workload~\cite{
cloudscaling-3}. 

With the possibility of extended cloud environments, as envisioned in fog computing that integrates a large volume of limited hardware resource edge nodes, the above challenges will need to be addressed in this new context~\cite{edgecomputing-00,fogcomputing-00}. The techniques that are employed for cloud resource management are scalable in a data center and even multiple data centers, but assume the concentration of resources. Edge specific characteristics, such as offering a service from a resource distant from the cloud, but closer to a user device, and computing on resource limited nodes will need to be accounted for. 
{\color{black} Although resource negotiators \cite{revision_yarn,revision_mesos} have been proved efficient in distributed clusters, they are specifically designed for large jobs, such as Hadoop or MapReduce and merely focus on pre-deployment resource provisioning.} Simply applying existing cloud-based techniques in the edge context will not be feasible since resource management on the edge will need to be lightweight (consuming minimal resources) {\color{black} and able to dynamically manage resources after deployment in order to support multiple tenants in resource deprived environments}. There is limited research exploring lightweight resource management techniques on the cloud~\cite{benchmarking-1} that could be directly applied in the context of the edge. However, further research in the edge context will be required to efficiently manage edge resources. In this paper, we set out to investigate techniques suited for the fog computing model, such that the QoS can be improved.    

{\color{black} Reference architectures for edge/fog computing have recently been released~\cite{revision_mecra,revision_openfogra}, which highlight the necessary functional layers. These provide a high-level description instead of an implementation. However, in this paper we aim to realise our proposed resource management architecture by implementing ENORM for a real world application.}

In the fog computing model, edge nodes are currently used in the following four ways to execute workloads. Firstly, in an aggregating model, in which data from multiple devices are collected by an edge node for pre-processing and filtering~\cite{
edgeaggregation-7} tasks. This model is conducive in sensor-based environments, such as wireless sensor networks~\cite{
edgeaggregation-6} and the upcoming Internet-of-Things (IoT)~\cite{
edgeaggregation-8}. 
An application gathers data from multiple devices or sensors and filters them on the edge node without routing the entire data to the cloud on which the application resides~\cite{
edgeaggregation-2}. 

Secondly, a sharing model is employed in which user devices, such as smartphones, tablets and laptops, that volunteer in a mobile cloud offer their spare computational cycles for executing a peer's workload~\cite{edgesharing-2}.
This model has been employed in video transcribing and face recognition use-cases~\cite{edgesharing-7}. The obvious disadvantage of this model is that it can only cover devices in a small region such as a shopping mall~\cite{edgesharing-4} or buses/trains~\cite{edgesharing-6} and is known for unstable computing due to the mobility of devices~\cite{
edgesharing-5}.

Thirdly, an offloading model can be employed in which workloads can be transferred from a user device to an edge node~\cite{
offloadd2e-6}. There is research highlighting the benefit of such an offloading model for workloads requiring numerical operations~\cite{offloadd2e-4}, for face recognition applications~\cite{offloading-1} and online games~\cite{ offloadd2e-8}. This is usually done in the context of a mobile edge cloud~\cite{offloadd2e-7} or cloudlets~\cite{offloadd2e-1}. 

The above three models cannot fully leverage the benefit of the edge for improving the QoS of a cloud application. This is because in the aggregating model, the server resides on the cloud and the edge node is only employed for pre-processing data which does not reduce the round-trip latency for a user. In the sharing model, given that peer nodes owned by individual users may not be able to offer continuous edge services, they cannot be employed to host application servers. In the offloading model from the user device to the edge, there is a limited case for complementing the computing requirements of user devices. Applications that are hosted as services require frequent communication between multiple users and the server. To improve the QoS, the frequency of communication between the users and the cloud will need to be minimised to improve the response time. This will be possible if computing is closer to the user, which is not supported in the above models. Hence, in this paper, we consider an alternate execution model of offloading workloads from the cloud server to an edge node.  

Current research in offloading workloads from a cloud server to an edge node focuses on caching~\cite{offloadc2e-4}, context-aware web browsing~\cite{offloadc2e-1} and video pre-processing~\cite{offloadc2e-7}. There is minimal research addressing resource management. Therefore, in this paper, we set out to address the three challenging problems in resource management when workloads are offloaded from the cloud to the edge. The first problem is partitioning a cloud workload for an edge node. In the context of edge nodes, there is limited research on partitioning servers suitably~\cite{chun2011clonecloud}, instead of deploying duplicate servers~\cite{offloadc2e-3}. The second problem is provisioning edge nodes and facilitating communication between the edge and the cloud servers. Existing edge-based research focuses on harnessing cloudlets without leveraging computing on nodes between the cloud and the user device along the data path~\cite{offloadc2e-5}. 
{\color{black} FocusStack is recently proposed for the discovery of edge nodes using geographic addressing in order to assist workload deployments~\cite{revision_focusstack}. Our paper on the other hand take the next step after discovery for managing resources by assuming that they have already been discovered using such approaches. Additionally, we consider a multi-tenant environment that is not considered in FocusStack. ParaDrop is designed to support multi-tenancy in Wi-Fi Access Points, with a cloud-based management to orchestrate applications across multiple edge nodes~\cite{revision_paradrop}. However, it only considers using the edge nodes as a replacement to clouds, which is fundamentally different to our approach in which both the cloud and edge are employed for improving the QoS of an application. Additionally, we have tested our approach on a real application.}
The third problem is dynamic management of resources on the edge to meet service objectives. There is auto-scaling research in the context of cloud VMs~\cite{autoscaleContainerInCloud}, but cannot be directly applied to the edge. Our research addresses this challenge in the context of containers. 
{\color{black} Techniques for auto resource provisioning and on-demand consumption of edge resources are proposed for IoT applications~\cite{revision_middleware} but do not support post-deployment resource management for multi-tenant environment.}

\section{Conclusions}
\label{sec:conclusions}
As more and more devices get added to the Internet, computing on nodes, such as routers, base stations and switches, at the edge of the network will need to be tapped into. This is the vision of fog computing. However, to realise this vision a number of problems will need to be solved. In this paper, we focused on the problems related to resource management. Existing resource management frameworks in distributed computing are suitable in the context of clouds and clusters, but do not integrate the edge of the network.

In this paper we have presented ENORM, an \textit{E}dge \textit{NO}de \textit{R}esource \textit{M}anagement framework that integrates edge nodes in the computing ecosystem to realise fog computing. ENORM addresses the resource management problems of provisioning edge nodes for cloud applications, deploying workloads on provisioned edge nodes, and dynamic resource allocation on edge nodes. To this end, we proposed a provisioning mechanism that considers handshaking, deployment of workloads and termination of edge services. Additionally, an auto-scaling mechanism to dynamically manage edge resources is developed. 
{\color{black} The provisioning and auto-scaling mechanisms are simple implementations {\color{black} based on linear search algorithm} given that edge nodes are resource constrained environments.}

The feasibility of ENORM was tested on a Pok\'eMon Go-like online game use-case. Experimentally it was noted that there are significant benefits in improving the QoS for a large number of users in a given location for the fog computing based use-case employing ENORM. When compared to a cloud-only model, the application latency is reduced between 20\%-80\%. Similarly, the data traffic and the communication frequency between the edge node and the cloud server are both reduced up to~{\color{black}95\%}. 

\subsection{\color{black}Limitations}
{\color{black}One shortcoming of the auto-scaler we have proposed is that it only considers static priorities of applications that are set by the cloud manager instead of dynamic priorities which are more realistic in real-world settings. The priority of an application may need to change when more users subscribe to the application. This will need to be considered with business models that make edge resources publicly available. Immediate efforts will be made to investigate this.}


{\color{black}The benefit of using fog computing for massively geo-distributed applications across data centers may not be obvious since the reduction in latency may not be significant to motivate the use of the edge of the network. However, the number of data centers is less likely to grow at the same rate as the number of devices since they consume lots of power and global network bandwidth. The edge of the network may be used to improve the QoS of applications.}

\subsection{\color{black}Future Work}
{\color{black} Currently, ENORM is suited for both single and multiple edge nodes environment. However, migrating applications between nodes is not considered in this paper. This will require synchronisation between edge nodes and suitable techniques for migrating or handing over a service onto another node. This will be investigated in the future.}

{\color{black} The current auto-scaling mechanism employed in ENORM adds a unit of both CPU and memory resources since it reduces the application latency of our use case. Other use cases may need to consider more resources (for example I/O) or a different combination of resources.} 
{\color{black}In addition, ENORM's performance may be affected by the granularity of the unit value. Our hypothesis is that when the unit value is highly fine-grained, the effect on the QoS of an application by scaling resources may be less obvious. A more coarse-grained value could result in an over provision resources required by the container. The optimal value for the unit value given an application may vary and will need to be explored through experiments. Determining a dynamic value by using heuristics may further optimise a static value that is used throughout the execution of an application. This will be investigated in the future.}

\section*{Acknowledgments}
The authors acknowledge the SFI-DEL grant (14/IA/2474).

\ifCLASSOPTIONcaptionsoff
  \newpage
\fi

\bibliographystyle{IEEEtran}  
\bibliography{references}

\begin{IEEEbiography}
[{\includegraphics[width=1in,height=1.25in,clip,keepaspectratio]{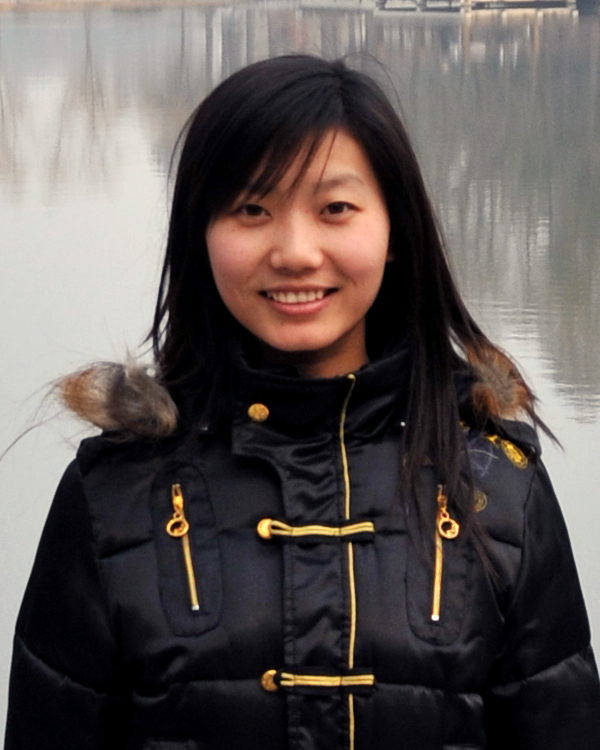}}]
{Nan Wang}
is a PhD student in Computer Science at Queen's University Belfast, UK. She obtained MRes in Web Science and Big Data Analytics (2015) from the University College London and MSc in Management and Information Technology (2014) from the University of St Andrews. She obtained her undergraduate degree from Beijing Jiaotong University, China. Nan's research interest is in fog computing. More information is available from \url{http://nwang03.public.cs.qub.ac.uk}.
\end{IEEEbiography}

\begin{IEEEbiography}
[{\includegraphics[width=1in,height=1.25in,clip,keepaspectratio]{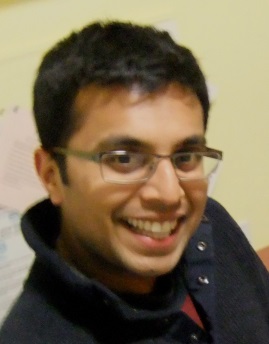}}]
{Blesson Varghese}
is a Lecturer in Computer Science at Queen's University Belfast and an Honorary Lecturer at the University of St Andrews. He obtained a PhD in Computer Science (2011) and MSc in Network Centred Computing (2008), both from the University of Reading, UK, on international scholarships. Blesson's interests are in developing and analysing novel parallel and distributed systems that leverage the edge of the network. More information is available from \url{www.blessonv.com}.
\end{IEEEbiography}

\begin{IEEEbiography}
[{\includegraphics[width=1in,height=1.25in,clip,keepaspectratio]{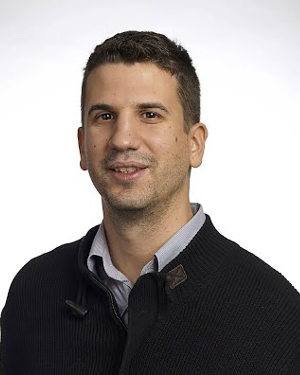}}]
{Michail Matthaiou}
is a Senior Lecturer at the ECIT Institute, Queen's University Belfast, UK. He obtained a PhD from the University of Edinburgh, UK in 2008 and received an MSc in Communication Systems and Signal Processing from the University of Bristol UK in 2005. Michail's interests are in signal processing for wireless communications, energy-efficient dense networks and fog computing.
\end{IEEEbiography}

\begin{IEEEbiography}
[{\includegraphics[width=1in,height=1.25in,clip,keepaspectratio]{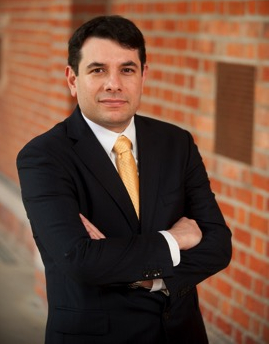}}]
{Dimitrios S. Nikolopoulos}
is Professor and Head of the School of Electronics, Electrical Engineering and Computer Science, at Queen's University of Belfast. He holds the Chair in High Performance and Distributed Computing. His research explores scalable computing systems for data-driven applications and new computing paradigms at the limits of performance, power and reliability. More information is available from \url{http://www.cs.qub.ac.uk/~D.Nikolopoulos/}.
\end{IEEEbiography}




\end{document}